\documentclass[%
reprint,
 superscriptaddress,
 amsmath,amssymb,
 aps,
pra,
floatfix,
noeprint
]{revtex4-2}

\usepackage{graphicx}
\usepackage{dcolumn}
\usepackage{bm}
\usepackage{amssymb}
\usepackage{epstopdf}
\usepackage{textcomp}
\usepackage{enumerate}
\usepackage{epsfig,amsmath}
\usepackage{subfigure}
\usepackage[colorlinks]{hyperref}
\usepackage{color}
\usepackage[usenames,dvipsnames]{xcolor}
\usepackage{float}
\usepackage{wasysym}
\usepackage{MnSymbol}
\usepackage{tabularx}
\usepackage{soul}
\usepackage{upgreek}

\begin{document}
\title{Strong magnetoresistance in a graphene Corbino disk at low magnetic fields}

\author{Masahiro Kamada}
\affiliation{Low Temperature Laboratory, Department of Applied Physics, Aalto University School of Science,
P.O. Box 15100, 00076 Aalto, Finland}

\author{Vanessa Gall}
\affiliation{\mbox{Institute for Quantum Materials and Technologies, Karlsruhe Institute of Technology, 76021 Karlsruhe, Germany}}
\affiliation{\mbox{Institut f\"ur Theorie der Kondensierten Materie, Karlsruhe Institute of Technology, 76128 Karlsruhe, Germany}}

\author{Jayanta Sarkar}%
\affiliation{Low Temperature Laboratory, Department of Applied Physics, Aalto University School of Science,
P.O. Box 15100, 00076 Aalto, Finland}

\author{\mbox{Manohar Kumar}}
\affiliation{Low Temperature Laboratory, Department of Applied Physics, Aalto University School of Science,
P.O. Box 15100, 00076 Aalto, Finland}
\affiliation{QTF Centre of Excellence, Department of Applied Physics, Aalto University, 
P.O. Box 15100, 00076 Aalto, 
Finland}
\author{Antti Laitinen}
\affiliation{Low Temperature Laboratory, Department of Applied Physics, Aalto University School of Science,
P.O. Box 15100, 00076 Aalto, Finland}

\author{Igor Gornyi}
\affiliation{\mbox{Institute for Quantum Materials and Technologies, Karlsruhe Institute of Technology, 76021 Karlsruhe, Germany}}
\affiliation{\mbox{Institut f\"ur Theorie der Kondensierten Materie, Karlsruhe Institute of Technology, 76128 Karlsruhe, Germany}}
\affiliation{Ioffe Institute, 194021 St.~Petersburg, Russia}

\author{Pertti Hakonen}
\email{pertti.hakonen@aalto.fi}
\affiliation{Low Temperature Laboratory, Department of Applied Physics, Aalto University School of Science,
P.O. Box 15100, 00076 Aalto, Finland}
\affiliation{QTF Centre of Excellence, Department of Applied Physics, Aalto University, 
P.O. Box 15100, 00076 Aalto, 
Finland}

\date{\today}

\begin{abstract}
We have measured magnetoresistance of suspended graphene in the Corbino geometry at magnetic fields up to $B=0.15$\,T, i.e., in a regime uninfluenced by Shubnikov-de Haas oscillations. The low-temperature relative magnetotoresistance  $[R(B)-R(0)]/R(0)$ amounts to $4000 B^2\% $ at the Dirac point ($B$ in Tesla), with a quite weak temperature dependence below 30\,K. A decrease in the relative magnetoresistance by a factor of two is found when charge carrier density is increased to $|n| \simeq 3 \times 10^{-10}$ cm$^{-2}$. 
The gate dependence of the magnetoresistance allows us to characterize the role of scattering on long-range (Coulomb impurities, ripples) and short-range potential, as well as to separate the bulk resistance from the contact one. Furthermore, we find a shift in the position of the charge neutrality point with increasing magnetic field, which suggests that magnetic field changes the screening of Coulomb impurities around the Dirac point. The current noise of our device  amounts to $10^{-23}$ A$^2$/$\sqrt{\textrm{Hz}}$ at 1\,kHz at 4\,K, which corresponds to a magnetic field sensitivity of 60 nT/$\sqrt{\textrm{Hz}}$  in a background field of 0.15\,T.


\end{abstract}

\maketitle

\section{Introduction}
Besides extraordinary physical characteristics, graphene exhibits superb electrical transport properties  \cite{Novoselov2004, CastroNeto2009}. Charge carrier conduction in monolayer graphene can display ballistic behavior over several microns, though the mean free path is often limited by Coulomb scattering and short-range scatterers \cite{Peres2010}. Typically Coulomb scatterers, embedded in the substrate or caused by fabrication residues, dominate the transport and short-range scattering becomes important only at large carrier densities. Using freely suspended graphene flakes and current annealing \cite{Bolotin2008}, however, impurity scattering can be minimized and intrinsic properties of graphene can be reached. Many of the basic transport properties of graphene have been revealed using suspended devices. Suspended graphene in Corbino disk geometry, for example,  has turned out to be valuable in sensitive investigations of fractional quantum Hall states in graphene \cite{Kumar2018}.

Magnetoconductance is a powerful tool for studying basic quantum transport in monolayer graphene \cite{Novoselov2005,Zhang2005,CastroNeto2009,Zhou2020}. Typically, magnetoconductance at low magnetic fields and low temperatures is governed by impurity scattering which leads to quantum corrections to the conductance and universal conductance fluctuation contributions \cite{ DasSarma2011}. For classical magnetoresistance, both linear and quadratic behavior is expected according to effective medium theory \cite{Tiwari2009,Ping2014}. Owing to strong demand for magnetic field sensors based on magnetotransport, various ways to generate large magnetoresistance in graphene have been developed in monolayer \cite{Zhou2020, Cho2008,Friedman2010,Bai2010,Gopinadhan2013,Wang2014,Chuang2016,Laitinen2018,Chuang2018a,Song2019,Hu2020} and multilayer graphene \cite{Gopinadhan2015a,Li2017}. 
In this work, we demonstrate that intrinsic behavior of suspended graphene in the Corbino ring geometry, already as such, yields a huge magnetoresistance.
This magnetoresistance can serve as an efficient tool for sample characterization as well as it can be used for sensing purposes. Our work demonstrates a magnetic field sensitivity on the order of 60 nT/$\sqrt{\textrm{Hz}}$ at 4\,K in a background field of 0.15\,T.


In the Hall bar geometry, both the transverse and longitudinal bulk conductivities -- $\sigma_{xy}(B)$ and $\sigma_{xx}(B)$, respectively -- determine the resistivity $\rho_{xx}(B)$. As a result, $\rho_{xx}(B)$ turns out to be independent of the applied magnetic field $B$ in the simplest one-band model. The Corbino-ring measurement setting for magnetoresistance is special already for this simplest case, because the Hall conductivity $\sigma_{xy}$ (or the Hall voltage) drops out from the resistivity. The latter is then obtained just as the inverse of the longitudinal conductivity $\rho_{xx}(B)=1/\sigma_{xx}(B)$. In a way, a Corbino disk is equivalent to an infinitely wide sample, in which the effect of the side walls can be neglected. The magnetoresistance for a generic anisotropic Corbino samples is calculated in Ref.~\cite{Nomokonov2019}; in the isotropic case, relevant to our setup, the resistance of the Corbino sample is expressed through to the bulk resistivity as 
\begin{equation}
    R(B)=\frac{1}{2\pi}\ \rho_{xx}(B)\ \ln\frac{r_\text{out}}{r_\text{in}}.
    \label{Corbino-resistance}
\end{equation}
Here $r_\text{in}$ and $r_\text{out}$ are, respectively, the inner and outer radii of the disk.
The logarithmic geometrical factor in Eq. (\ref{Corbino-resistance}) reflects the total current conservation in the Corbino disk. 

Bulk magnetoresistance measured in disordered graphene in the Hall-bar geometry may display complex magnetic field dependence \cite{Ando2002,Sachdev2008, Jobst2012, Alekseev2013}; for example, indications of $\sqrt{B}$ dependence at small fields have been reported \cite{Vasileva2012,Vasileva2019}. Our results, on the contrary, display a strong parabolic ($B^2$) magnetoresistance for arbitrary disorder, while only small corrections to the $B^2$ dependence are found for magnetic fields up to 0.1 T.

\section{Sample fabrication and characterization}

\begin{figure}[b]
	\centering
	\includegraphics[width=0.95\linewidth]{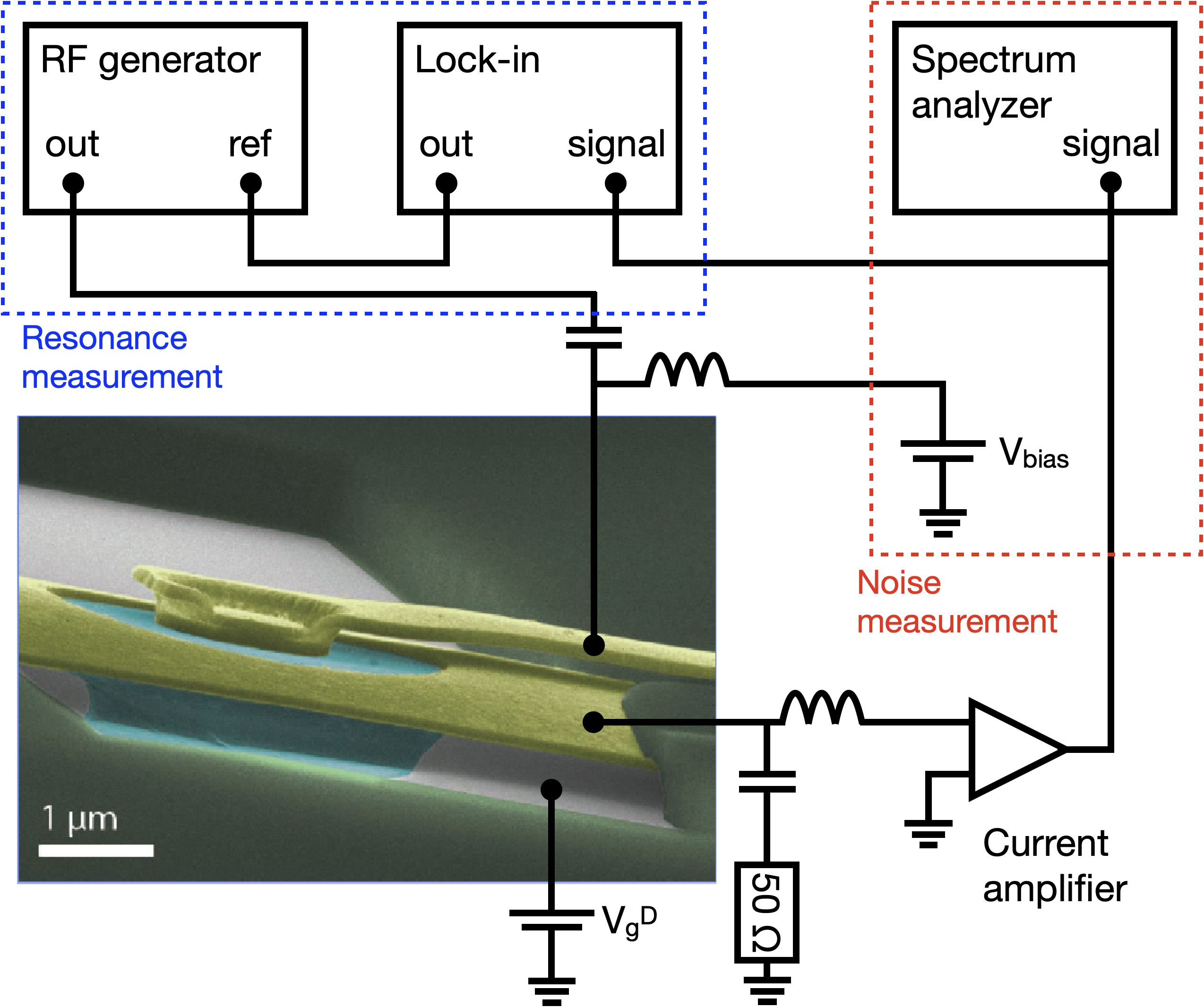}
	\caption{Schematics of our measurement configuration and a scanning electron microscope image of a suspended graphene Corbino disk (green part in the center) with 4.5 $\mu$m diameter outer Au contact and 1.8 $\mu$m diameter inner Au contact.   \label{DR_R_vs_B2} 
}
\end{figure}

Our graphene samples were fabricated using a technique based on lift off resist (LOR) sacrificial layer \cite{Tombros2011}; details of the employed process can be found in Ref. \cite{Kumar2018}. The current annealing of the samples at low temperature before measurements guaranteed a high maximal field-effect mobility $\mu_{\text{FE}}^{\text{max}} \simeq 1-2\times 10^5$ cm$^2$/Vs. Most of the data was measured on a Corbino disk with inner and outer radii of 0.9 and 2.25 $\mu$m, respectively. A false-color scanning electron microscope picture of a Corbino sample is displayed in Fig. \ref{DR_R_vs_B2}, together with its connections to the employed measurement devices. The gate capacitance $C_g = 1.5 \times 10^{-5}$ F/m$^2$ was obtained using the Landau level fan diagram \cite{Kumar2018}.

In our experiments, we employed standard voltage-biased measurements and recorded current through the sample over the bias range of $1-13$ mV. These rather large voltages were chosen in order to measure the low-frequency noise spectra simultaneously and the resistance noise of our samples was in the range of $\delta R/R \lesssim 2 \times 10^{-5}$ at 1 Hz.  The current was amplified using a transimpedance amplifier at gain $10^5$\,V/A. Bias-T components facilitated insertion of rf-signals to the sample. For details of the measurement system, see Ref.~\cite{Laitinen2018c}. In the measurements, both positive and negative bias voltages $V$ were used. A weak $V$ dependence was removed by extrapolating data at $V<0$ and at $V>0$ down to zero bias: the two extrapolations differed less than 3\%.

\begin{figure}[t]
	\centering
	\includegraphics[width=.95\linewidth]{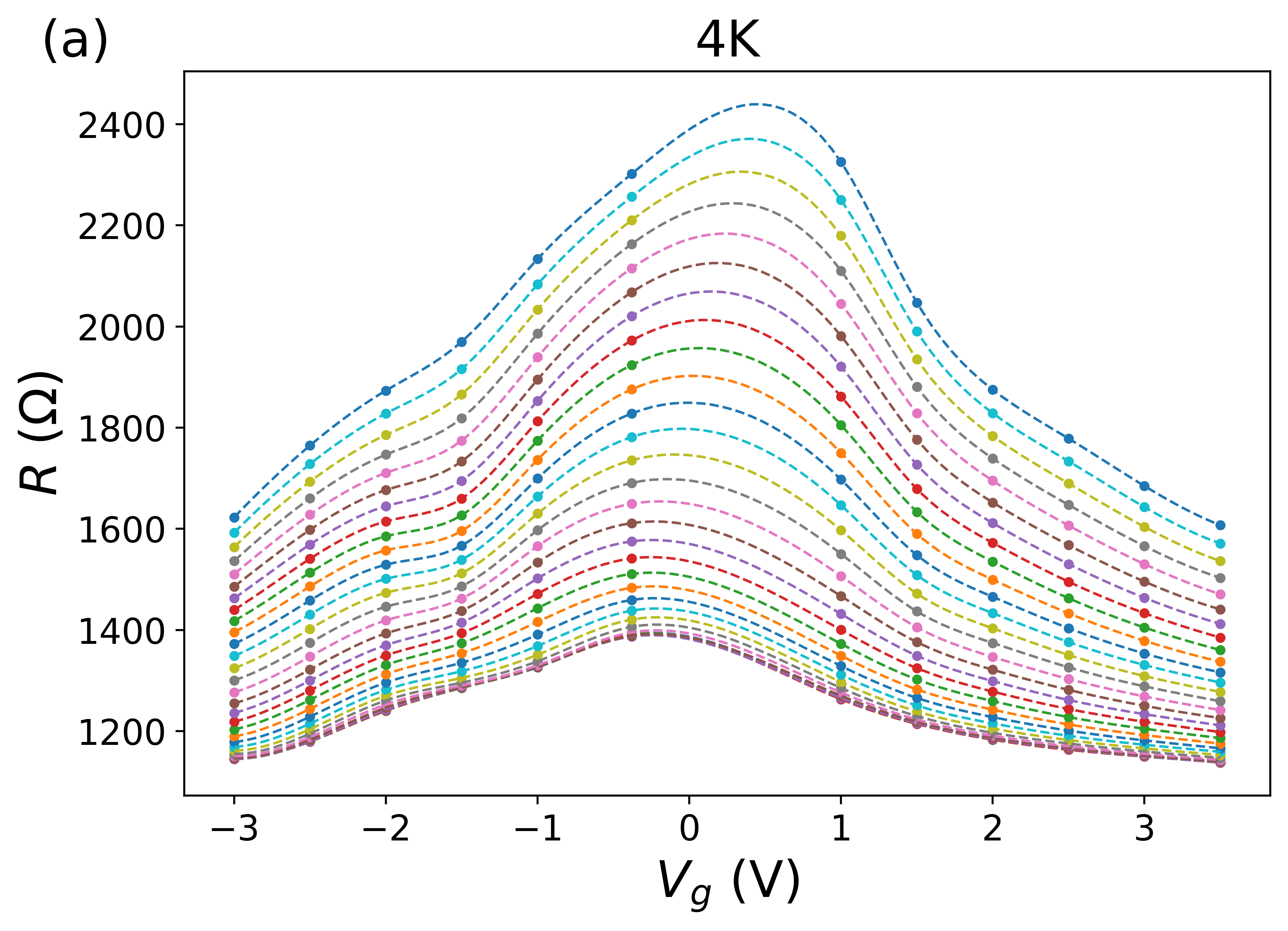}
	\includegraphics[width=.95\linewidth]{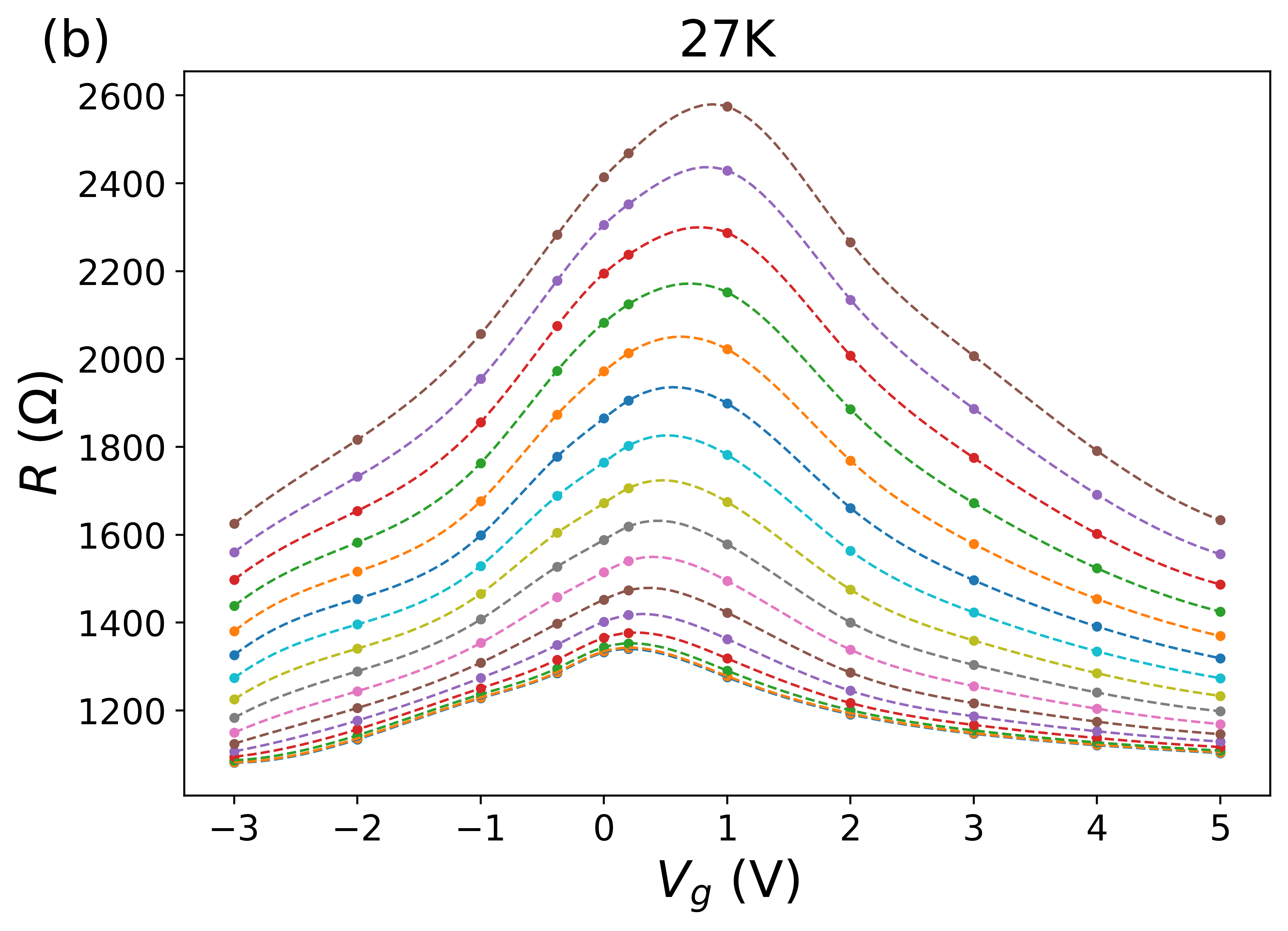}
\caption{
Resistance vs. $V_g$ at various magnetic fields. (a) $4$\,K for magnetic fields from 0 to 0.15 T with 0.005 T step (from bottom to top); the broken curves are cubic spline interpolations to the data. (b)  $27$\,K for magnetic fields from 0 to 0.15 T with 0.01 T step (from bottom to top); the broken curves are cubic spline interpolations. \label{fig:bothRes}}
\end{figure}

First, we characterized the gate voltage dependence of the resistance $R(V_g)$ of our sample at $B=0$ (see lowest curves in Fig. \ref{fig:bothRes}) at two temperatures, $T=4$\,K and $T=27$\,K. The offset of the Dirac point from zero amounted to $V_g^{D}(0) \simeq 0.2$\,V at $T=27$\,K, and the corresponding residual charge density was found to be $n_0 \simeq 8 \times 10^{9}$ cm$^{-2}$. For $T=4$\,K, the zero-$B$ offset of the Dirac point had the opposite sign: $V_g^{D}(0) \simeq -0.2$\,V. The temperature dependence of the offset (and, in particular, the change of its sign) 
can be related to different concentration of {(quasi)resonant} adsorbed atoms, see below. 

Figure \ref{fig:bothRes} includes also our data at small magnetic fields, $B<0.15$\,T. For both temperature values, the resistance grows strongly with $B$ and, simultaneously, there appears a shift of the maximum value to slightly higher gate voltage $V_g^{D}(B)$. The data points were joined using cubic spline fits in order to make observed changes with $B$ easier to distinguish. According to the spline fits, the shift of the Dirac point for $T=27$\,K amounts to $\Delta V_g = V_g^{D}(B)-V_g^{D}(0)= 5 B$\,V, where $B$ is measured in Tesla. Even stronger shift is seen for $T=4$\,K.

The observed shift of the Dirac point with increasing $B$ can be attributed to changes in screening of charged impurities in a magnetic field near charge neutrality (see Appendix \ref{sec:Diracshift} for more details). In this case, the total charge density induced on the membrane is less at the Dirac point at finite $B$, which would indicate generation of more positive charge on graphene by the screening. Thus the screening should take place by positive carriers and the impurities are negatively charged. In addition, there was a slow shift of the Dirac point position towards positive $V_g$ over time (on the order of 0.2\,V in one month). It is worth noting that the effect of magnetic field on screening is suppressed at higher densities and higher temperatures. This is in line with the stronger shift of the resistance maximum at $T=4$\,K. 

One more peculiarity seen in the resistance curves plotted over the gate voltage is a feature close to $-2$\,V for both temperatures shown in Fig.  \ref{fig:bothRes}. The resistance around this voltage is somewhat enhanced compared to the resistance away from this voltage. There is no comparable feature at the electron side of the resistance curve and the feature is stronger for lower temperatures. We attribute this feature to a broadened resonance level associated with adsorbed local impurities. When the chemical potential is moved by the gate voltage into the vicinity of this quasi-resonance, the scattering amplitude for such impurities is enhanced, leading to shorter transport scattering time and, hence, to the increase in resistance. At the same time, the broadening of this resonance is sufficiently strong to avoid truly resonant scattering (as, e.g., in the case of vacancies); in contrast to infinitely strong impurities (vacancies), the position of the quasi-resonance is shifted away from the Dirac point.
Away from the resonant energy, these impurities produce weak short-ranged disorder.

For higher temperatures, some adsorbed dirt is thermally removed form the sample, leading to a less pronounced feature. This suggests that the role of scattering off short-range disorder at higher temperatures could be decreased. Below, based on the analysis of the magnetoresistance curves, we will discuss this issue in more detail. The dependence of the concentration of local quasi-resonant impurities
on temperature can also explain the $T$ dependence of zero-$B$ shift of the Dirac point mentioned above. Indeed, at higher temperature ($T=27$\,K in Fig.~\ref{fig:bothRes}), the shift of the chemical potential is smaller, which is consistent with the above picture of lower concentration of adsorbed impurities.

\begin{figure}[t]
	\centering
	\includegraphics[width=\linewidth]{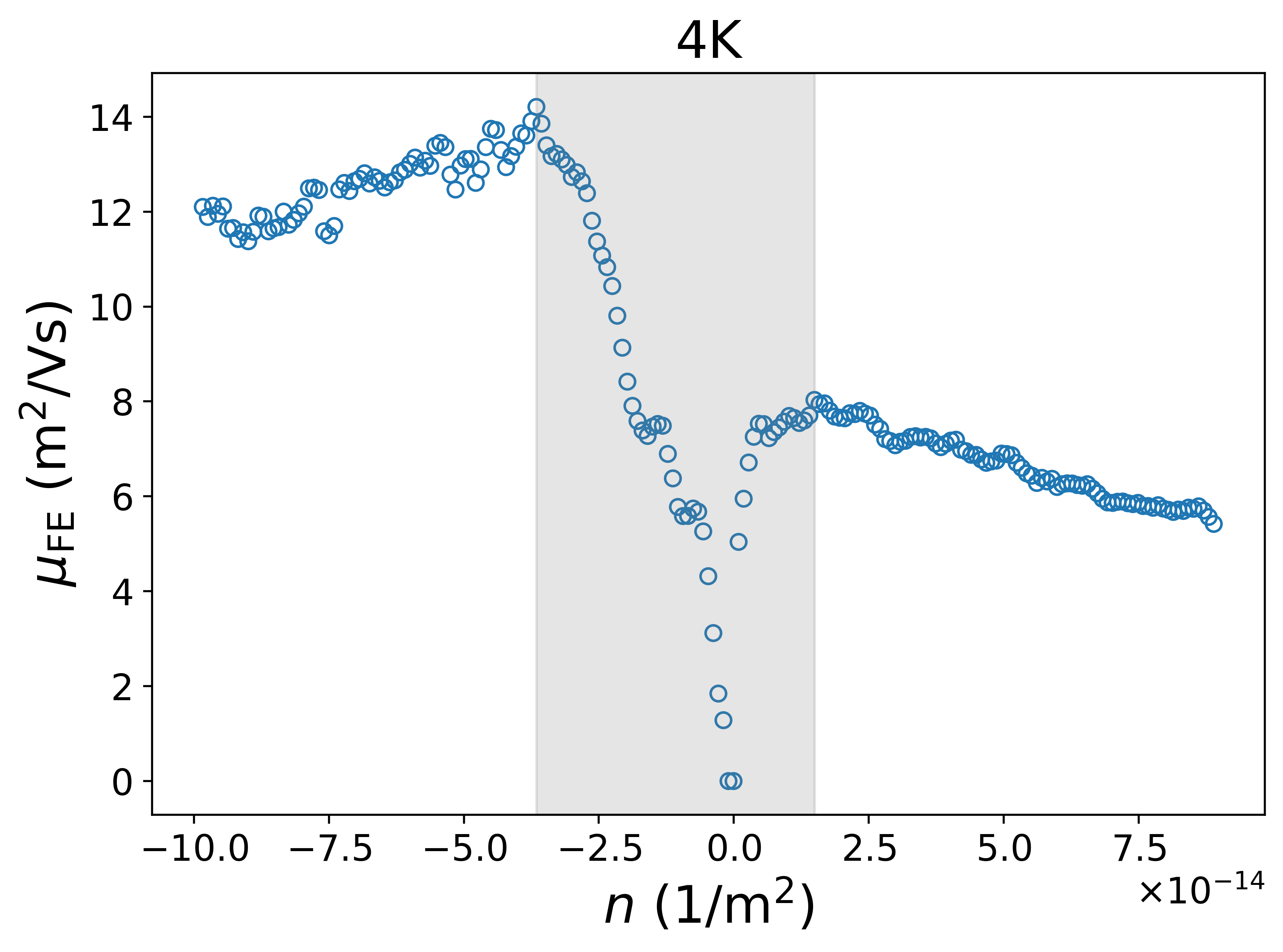}
\caption{
Field effect mobility  $\mu_{\text{FE}}$ at $T=4$\,K determined  from the measured differential resistance $dV/dI$ at $B=0$. The Dirac point shift $V_g^D(0)$ has been subtracted off from the gate voltage before calculating the charge carrier density $n$. The dip in $\mu_{\text{FE}}$ around $n=0$, indicated by a grey shadow, corresponds to the density range governed by disorder broadening of the Dirac point. The extend of this range is consistent with the value $n_*$ given in Table~\ref{tab:parameters} \label{fig:FE}}
\end{figure}

In Sec. \ref{sec:analysis}, we deduce charge carrier mobility from the measured geometric magnetoresistance. For comparison, we display in Fig. \ref{fig:FE} the field-effect mobility defined by $\mu_{\text{FE}}=e^{-1}d\sigma_{xx}/dn$, obtained from measurements of differential resistance  $R(V_g)$ at the end of the experiments. These data measured at $T=4$\,K indicate that, for our sample at a tiny bias voltage, the maximum mobility for holes $\mu_{\text{FE}}\simeq 13$\,m$^2$/Vs is clearly larger than that for electrons $\mu_{\text{FE}} \simeq  7$\,m$^2$/Vs. However, at the employed bias voltages $1 \dots 10$\,mV, the influence of the pn interface at positive gate voltages appears to be reduced and $\mu_{\text{FE}}$ for electrons and holes becomes almost equal. 

\section{Theoretical background}

Before analyzing the obtained magnetoresistance data,
we present in this Section the basic facts about disorder-dominated transport in graphene (for the hydrodynamic---collision-dominated---transport, higher temperatures are typically required than those in our experiment, $50$\,K $<T<150$\,K, while phonons become important at yet higher $T$ \cite{Gornyi2012}).
Distinct from conventional electron gases in 2D semiconductor heterostructures, graphene displays a linear energy dispersion relation of the carriers,
$ \varepsilon_{\mathrm{k}} = \pm v \hbar k$.
This leads to the linear-in-energy
density of states in clean graphene: 
\begin{align}
\nu_0 (\varepsilon)=
\frac{N|\varepsilon|} { 2 \pi v_F^2 \hbar^2},
\label{dos}
\end{align}
where $v_F=10^6~\text{m/s}$ is the Fermi velocity, $\varepsilon$ is the energy counted from the Dirac point,  and  $N=4$ is the degeneracy due to spin and valley degrees of freedom. Following from this density of states, the charge density of carriers is given by
 \begin{align}
n = N\frac{\varepsilon_F^2}{4\pi\hbar^2v_F^2},\label{eq:density}
 \end{align} 
where $\varepsilon_F$ is the Fermi energy. 
A consequence of the linear dispersion relation is that
the cyclotron frequency $\omega_{c}$ becomes energy-dependent \cite{CastroNeto2009} :
\begin{align}
\omega_{c}(\varepsilon)=\frac{eB}{m_c(\varepsilon)}=\frac{\hbar}{m_c(\varepsilon)\ell_B^2},
\label{omega-c}
\end{align}
where $B$ is the magnetic field and $m_c(\varepsilon)={\varepsilon}/{v_F^2}$ is the cyclotron mass, also dependent on the energy, and where we have defined a magnetic length by $\ell_B= \sqrt{\hbar/eB}$. 

The linear dispersion relation also influences the scattering and relaxation rates of charge carriers.
We assume that the major contributions to resistance arise from short-range scattering (s) and Coulomb scattering (C) and adopt this mixed-disorder model \cite{Hwang2007} to describe the magnetoresistance. For the interpretation of the effect of charged scatterers, one should keep in mind, that graphene is not perfectly flat, but has ripples \cite{Meyer2007, Lau2012, Gornyi2015}. The scattering off ripples is similar to that for charged impurities \cite{Gornyi2012}, so that their contribution is effectively included in our treatment of Coulomb scattering.

We first address scattering on the short-range impurity potential.
We limit our discussion to not too high carrier densities for which the length scale $d$ for random potential variations is smaller than the carriers wave-length and larger than the interatomic spacing $a$, i.e.,  $a \ll d \ll \lambda$. For such disorder the intervalley scattering can be disregarded and the  quantum $\tau_q$ and transport $\tau_\text{tr}$ scattering times can be estimated using Fermi's golden rule \cite{ShonAndo1998,CastroNeto2009}:
 \begin{equation}
\label{tau0} 
\tau_{{q} }^{s}(\varepsilon)=\frac{\hbar \gamma_s
}{|\varepsilon|} \: , \;\;\;\; 
\tau_{\text{tr} }^{s}
(\varepsilon)=2\tau_{{q}}^{s} (\varepsilon) \:,
\end{equation}
where 
\begin{equation}
    \gamma_s=\frac{2 \hbar ^2 v_F^2}{n_\text{imp}^{s} U_0^2},
\end{equation}
$n_\text{imp}^{s}$ is the concentration of short-range impurities, and  $U_0$ denotes the magnitude of the impurity potential. 
In what follows, we will characterize the strength of short-range disorder in the samples by the parameter $\gamma_s$ which is energy independent.
The difference between  $\tau_{\text{tr}}$ and  $\tau_q$ is caused by the weak scattering anisotropy which originates from the spinor nature of the wave functions. 

The scattering times for charged impurities can be brought to similar form (see Appendix \ref{sec:AppCharge}), although the effective parameter $\gamma_C$ is then, in general, a function of the energy and the Fermi energy, as well as temperature and magnetic field (through the corresponding dependence of screening of Coulomb impurities by charged carriers). Below, we mainly consider sufficiently low temperatures and magnetic fields, keeping only the energy dependence of $\gamma_C$: 
 \begin{equation}
\label{tau1}
\tau_{{q} }^{C}(\varepsilon)
=\frac{\hbar\gamma_C^{\prime}(\varepsilon,\varepsilon_F)}
{|\varepsilon|} \: , \;\;\;\; 
\tau_{\text{tr} }^{C}(\varepsilon)=2\frac{\hbar  \gamma_C(\varepsilon,\varepsilon_F)}
{|\varepsilon|} \:,
\end{equation}
Here $\gamma_C(\varepsilon,\varepsilon_F)$ and $\gamma_C^{\prime}(\varepsilon,\varepsilon_F)$ are functions of the effective coupling (graphene ``fine-structure constant'')
$\alpha=e^2/(\hbar v_F \epsilon_{\infty})$,  
with $\epsilon_{\infty}$ the background dielectric constant, see Appendix \ref{sec:AppCharge}.
In the absence of a screening environment and neglecting the renormalization of velocity by Coulomb interaction, the nominal value of this constant is $\alpha_0=2.2$. However, both the screening and renormalization \cite{CastroNeto2009} effects reduce this value.
For intermediate values of $\alpha$, the 
relation 
$\tau_{\text{tr} }^{C}(\varepsilon)\approx 2 \tau_{{q} }^{C}(\varepsilon)$, similar to Eq. (\ref{tau0}), holds, but it is no longer exact.

The  conductivity in zero  magnetic field in the presence of only short-range scatterers is given by the Drude formula: 
\begin{equation}
\sigma_{xx}^D  = \sigma_0 = \frac{2e^2\gamma_s}{\pi\hbar}. 
\label{Drude}
\end{equation}
By comparing this with the typical conductivity of high-quality graphene samples, we observe that $\gamma_s$ should be of the order of unity 
if impurities in high-mobility samples, such as ours,
were short-ranged (this estimate corresponds to measured quantum scattering time at $V_g = 10$\,V and scaled to our measurement regime).

An important parameter for the magnetoresistance is the product $\omega_{{c}} \tau_q^{s}$ which describes the broadening of Landau levels. Since for short-range scatterers both  $\omega_{{c}}$ and $\tau_q^{s}$ depend on energy [see Eqs.\eqref{omega-c} and \eqref{tau0}], the parameter
\begin{equation}
x=\omega_{{c}} \tau_q^{s}=\frac{\gamma_s\hbar^2 v_F^2}{\varepsilon^2 \ell_B^2},
\label{x}
\end{equation}
can be either small or large, depending on the energy $\varepsilon$ \cite{Alekseev2013}. 
For scattering on Coulomb impurities, the quantum scattering time decreases linearly with energy $\varepsilon \propto \sqrt{n}$, which means that the parameter $x=\omega_{{c}} \tau_q^{C}$ becomes energy independent. The parameter $x$ determines the dependence of  the density of states $\nu(\varepsilon)$ of disordered graphene on magnetic field. Since the transport scattering time $\tau_\text{tr}$ has the same energy dependence as $\tau_q$, the same parameter $x$ governs the quasiclassical bending of particle trajectories in magnetic field.   

The general result for the longitudinal conductivity $\sigma_{xx}(\varepsilon)$ is given by Eq.~(4.13) of Ref.~\cite{ShonAndo1998}. Introducing the relative density of states $\tilde{\nu}(\varepsilon)=\nu(\varepsilon)/\nu_0(\varepsilon)$, where $\nu_0(\varepsilon)$ is the zero-field density of states, we write the conductivity kernel (conductivity of particles at energy $\varepsilon$) in terms of $\tilde{\nu}(\varepsilon)$ as follows:
 \begin{equation}
\label{sigma_xx} 
\sigma_{xx}(\varepsilon)=\sigma_0
 \frac{\tilde{\nu}(\varepsilon)^2}{\tilde{\nu}(\varepsilon)^2+
 [\omega_{\mathrm{c}}(\varepsilon)\tau_\text{tr}(\varepsilon)]^2},
\end{equation}
where
\begin{align}
\sigma_0&=\frac{e^2 v^2}{2}
\tau_\text{tr}( \varepsilon_F)
\nu( \varepsilon_F)\equiv 
\frac{e^2\gamma N}{2\pi \hbar}. 
\label{eq:defGamma}
\end{align}
Here, we have introduced the dimensionless disorder strength $\gamma$ that has a meaning of a dimensionless conductance per spin per valley. For the short-range disorder, $\gamma=\gamma^s$. In the presence of both short-range and Coulomb scatterers, the total transport time is determined by
\begin{align}
\frac{1}{\tau_\text{tr}(\varepsilon)}
=\frac{1}{\tau_\text{tr}^{s}(\varepsilon)}
+\frac{1}{\tau_\text{tr}^{C}(\varepsilon)}, \label{eq:totaltransporttime}
\end{align} 
and $\gamma$ is related to the total transport scattering time, as given by Eq.~(\ref{eq:defGamma}).

The  finite-temperature conductivity is given by the kernel
(\ref{sigma_xx}) integrated with the derivative of the Fermi function $n_F(\varepsilon)$:
\begin{align}
\sigma_{xx}=\int_{-\infty}^{\infty}\mathrm{d}\varepsilon\left(-\frac{\partial n_F(\varepsilon)}{\partial\varepsilon}\right) \sigma_{xx}(\varepsilon) 
\label{eq:TDependence}
\end{align}
At zero temperature, 
the derivative gives the delta-function and the 
conductivity reduces to Eq.~(\ref{sigma_xx}) with $\varepsilon\to \varepsilon_F$. At zero magnetic field it is given by $\sigma_0$ from Eq.~(\ref{eq:defGamma}).
The temperature dependence of the Drude conductivity arises from the energy dependence in the kernel (\ref{sigma_xx}) when the thermal broadening of the delta-function is taken into account. At low temperatures,  $k_B T \ll \varepsilon_F$, the finite-$T$ corrections to the zero-$T$ result are small, and the measured conductivity is given approximately by 
$\sigma_{xx}\approx\sigma_{xx}(\varepsilon_F,T=0)$.
Under these conditions, if $\tilde{\nu}_0$ is independent of the magnetic field, the conductivity in a finite magnetic field can be written in the 
conventional Drude form: 
\begin{align}
    \sigma_{xx}(B) = \frac{en\mu_0}
{1 +(\mu_0 B)^{2}},
\end{align}
where $\mu_0$ is the mobility at $B=0$, i.e,
\begin{align}
    \mu_0=\frac{\sigma_{xx}(B=0)}{ne}.\label{defMobility}
\end{align}
As can be seen from the calculation of $\tilde{\nu}_0(\varepsilon)$ in Refs.~\cite{Ostrovsky2006,Alekseev2013}, corrections to the density of states arising from the finite magnetic field can be neglected, as long as $x\ll 1$, which we will show is true over a large range of data in the present experiment.

Comparison with Eq.~(\ref{sigma_xx}) shows that $\omega_c(\varepsilon) \tau_{\text{tr}}(\varepsilon)/\tilde{\nu}_0(\varepsilon)$
corresponds to $\mu_0B$. This means that the Drude conductivity is given by
\begin{align}
\sigma_{xx}\approx\frac{e^2\gamma N}{2\pi \hbar}\frac{1}{1+\left(\frac{2 e \gamma}{\pi\hbar n}\right)^2B^2}\label{eq:sigmaxxzeroT},
\end{align}
which has finite temperature corrections that are detailed in the Appendix \ref{sec:AppT}.
There are also temperature-dependent quantum corrections to the Drude conductivity, in particular, those arising from the electron-electron interaction (EEI), as discussed in Appendix \ref{sec:AppEEI}. 

For both short-range and Coulomb impurities, Eq.~(\ref{sigma_xx}) then yields a parabolic magnetoresistance in the Corbino geometry.
The Drude resistivity in the Corbino geometry 
takes a simple form
\begin{align}
\rho_{xx}(B)=\frac{1}{\sigma_0}
\left[1+(\mu_0 B)^2\right].
\label{eq:resistivity}
\end{align}
According to the Mathiessen rule, the inverse mobility can be written as a sum of the contributions of different momentum-relaxing scattering processes
\begin{align}
  \mu_0^{-1}&=\mu_C^{-1}+\mu_s^{-1},
\end{align}
 which yields in the zero-$T$ limit:
\begin{align}
\frac{1}{\mu_0}&=\frac{\pi \hbar n}{2 e \gamma}=\frac{\pi \hbar n}{2 e }
\left(\frac{1}{\gamma_s}
+\frac{1}{\gamma_C(\varepsilon=\varepsilon_F,\varepsilon_F)}\right)\label{eq:mobility},
\\
\gamma_C&(\varepsilon=\varepsilon_F,\varepsilon_F)
=\frac{n}{c(\alpha) n_\text{imp}^{C}}\label{eq:gamma_C} ,
\end{align}
with $c(\alpha)$ defined in Appendix \ref{sec:AppCharge}.
One sees that the contribution of Coulomb scatterers to the inverse mobility is density independent. On the other hand, the mobility governed by short-range impurities decreases with charge carrier density as $\mu_s \propto 1/n$. Thus, the density dependence of the total mobility allows one to characterize the role of short-range and Coulomb impurities in transport.  

Below a certain chemical potential or the corresponding density $n_*$, disorder-induced broadening smears the single-particles energy and the density of states saturates. This energy scale is given by the self-consistent equation for $\varepsilon$:
\begin{align}
\frac{\hbar}{\tau_q(\varepsilon_*)} \sim \varepsilon_* \label{eq:saturation}
\end{align}
For the mixed disorder model with $\gamma_s\gg 1$, we get for the corresponding density
\begin{align}
n_* \sim d(\alpha)n_\text{imp}^{C}, \label{eq:nstar}
\end{align}
where $d(\alpha)$ is given in Appendix~\ref{sec:AppCharge}. The value of $n_*$ depends on the density of charged impurities, fine-structure constant $\alpha$. We model this saturation effect by performing the replacement $n\rightarrow \sqrt{n^2+n_*^2}$, which effectively interpolates between $n$ at high densities and $n^*$ at the neutrality point, in all formulas, when used for plotting or fitting. In order to keep the notation clear, we do not explicitly write down this  replacement in the main text. Since this replacement is an approximate interpolation, it describes the behavior of the density of states (and other observables) at $n\sim n_*$ only qualitatively, see Appendix~\ref{sec:AppSat} for details. 
Nevertheless, this simple interpolation function allow us to confidently extract the system parameters, when the range of densities $n\gg n_*$ is included in the fit.

\section{Analysis of data and Results \label{sec:analysis}}

\begin{figure}[t!]
	\centering
	\includegraphics[width=.95\linewidth]{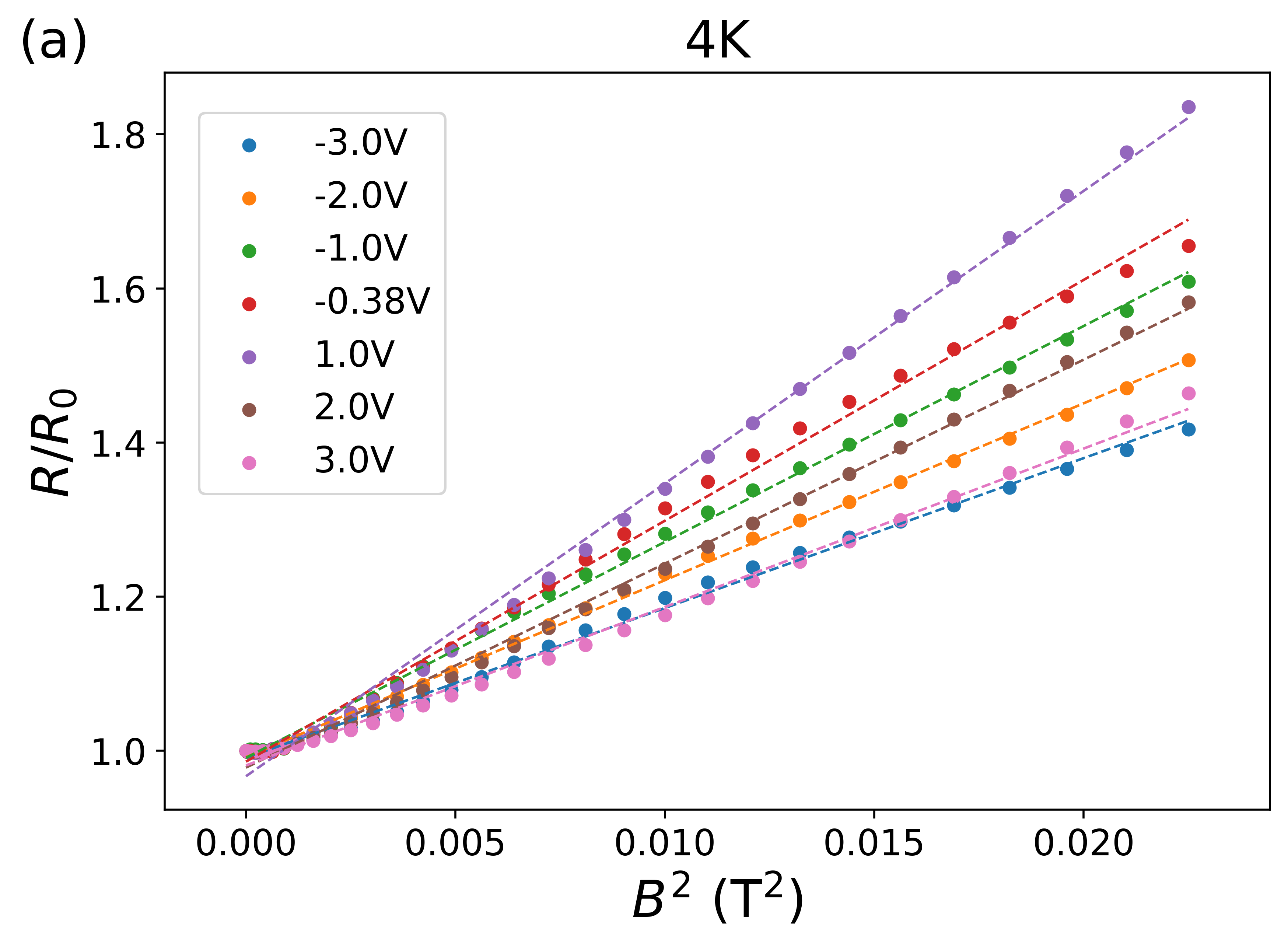}
	\includegraphics[width=.95\linewidth]{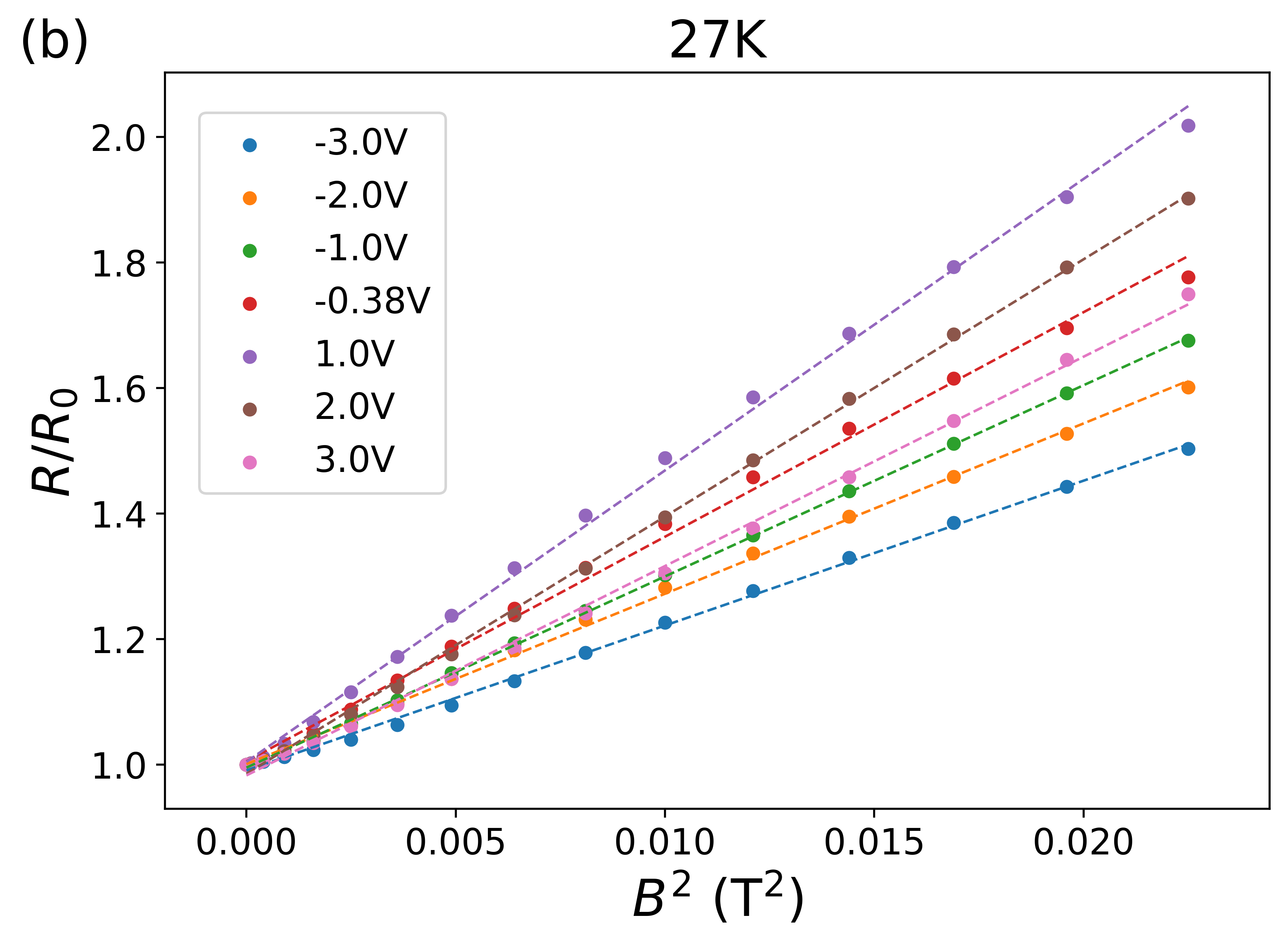}
\caption{(a) Scaled resistance $R/R(0)$ vs. $B^2$ at $T=4$\,K measured at various gate voltage values between $-3\dots +3$\,V ($0.8 \times 10^{9}$ cm$^{-2}$ $ < |n| < 3 \times 10^{10}$ cm$^{-2}$). The Dirac point corresponds to $V_g \simeq 1$\,V, the data at which is denoted by black symbols in the figure. The solid curves are
guides for the eye, emphasizing an overall parabolic magnetoresistance and slight deviations from parabolicity.
(b) Scaled resistance $R/R(0)$ vs. $B^2$ at $T=27$\,K measured at various gate voltage values between $-3\dots +3$\,V. The magnetoresistance at low fields grows faster at 27\,K than at 4\,K. 
	\label{MGresistance}}
\end{figure}

The relative magnetoresistance $\Delta  R(B)/R(0)=R(B)/R(0)-1$ of our sample at $B < 0.15$ T is illustrated in Fig.~\ref{MGresistance} which depicts the relative resistance $R(B)/R(0)$ as a function of $B^2$ measured at $T=4$\,K (Fig.~\ref{MGresistance}a) and at $T=27$\,K (Fig.~\ref{MGresistance}b). In both data sets, the magnetoresistance is found to be the strongest at the Dirac point, which is in agreement with the weakest effect of scattering when $|n|$ is smallest. 
Both data sets are influenced by the growing shift of the Dirac point $\Delta V_g$ as $B$ increases. 


The strength of the measured magnetoresistance depends only weakly on temperature up to 27\,K. However, when comparing the data at 4\,K and 27\,K, one observes that the $B^2$ dependence is followed better at 27\,K than at 4\,K in small magnetic fields. Qualitatively, this could be a signature of increased role of electron-electron scattering \cite{Jobst2012,Alekseev2013} and macroscopic inhomogeneities \cite{Ping2014,Vasileva2019,Vasileva2019a}.
In our suspended graphene sample such inhomogeneities can be due to static ripples.

The nature of scattering does not appear to play a large role in the measured magnetoresistance. The parabolic field dependence is followed for both Coulomb and short-range impurities in the range of parameters covered: $ |n| \sim 0.8 - 3 \times 10^{10}$ cm$^{-2}$ and $T=4-27$\,K.
In general, the $B^2$ dependence at small magnetic fields is more closely followed  in the 27\,K data. The 4\,K data displays deviations from $B^2$ behavior at $B < 30$ mT, which may be a sign of coherent behavior and quantum interference effects, either regular weak localization type or Corbino-geometry related as predicted for graphene in Ref.~\cite{Rycerz2010a}. 
At the largest magnetic fields around $0.1-0.15$ T, small deviations from $B^2$ dependence become obvious, in particular near the Dirac point. One can interpret this deviation as the onset of the Shubnikov - de Haas (SdH) oscillation regime in the sample \cite{Laitinen2018b} that corresponds to $x\sim 1$. 
 
On top of this one has to consider additional contribution $R_\text{cont}$ to the measured resistance related to the contact effects. These contributions are the resistance of the metal-graphene contacts and the interface resistance of the contact-doped graphene region. The former contribution is a microscopic material property, which we take to be constant. The latter, discussed in Ref. \cite{Laitinen2016}  is of the type of the $pn$-junction resistance. 
This contribution to the total resistance depends on the density of charge carriers in the bulk of the sample and is the main cause for the usually observed electron-hole asymmetry in transport measurements. 
In low magnetic fields the cyclotron radius is larger than the geometrical length scales characterizing the contact region and, hence, the overall contact resistance should not show a pronounced magnetic-field dependence.

The parabolic magnetoresistance is associated with the bulk contribution, whereas the total resistance includes the contact resistance: $R=R_\text{bulk}+R_\text{cont}$, where $R_\text{bulk}$ describes the disorder-induced bulk resistance and $R_\text{cont}$ the contact contribution. Since $R_\text{cont}$ depends on the gate voltage, the normalized magnetoresistance shown in Fig.~\ref{MGresistance} is not particularly convenient for extracting the density dependence of the mobility.
Indeed, the value of $R(B=0)$ in the denominator of the scaled magnetoresistance is not equal to $R_\text{bulk}(B=0)$ in front of the $B$-dependent term coming from Eq.~(\ref{eq:resistivity}), so that the coefficient in front of the $B^2$ term in the scaled magnetoresistance is not equal to $\mu_0^2$.

To overcome this complication caused by the contact resistance, we have employed a fitting function of the form
\begin{equation}\label{MGres}
R(B)-R(0)= M B^2 ,
\end{equation}
for the total resistance,
where we have a single fitting parameter $M$
fully determined by the properties of the bulk of the sample. According to Eq.~(\ref{eq:resistivity}) we have $M=R_\text{bulk}(0)\mu_0^2$, where $R_\text{bulk}(0)$ describes bulk resistance at zero field and $\mu_0$ is the mobility. We recall that $R_\text{bulk}(0)$ is different from the measured $R(0)$  because the latter includes the contact contribution.  

\begin{figure}[t!]
	\centering
	\includegraphics[width=.95\linewidth]{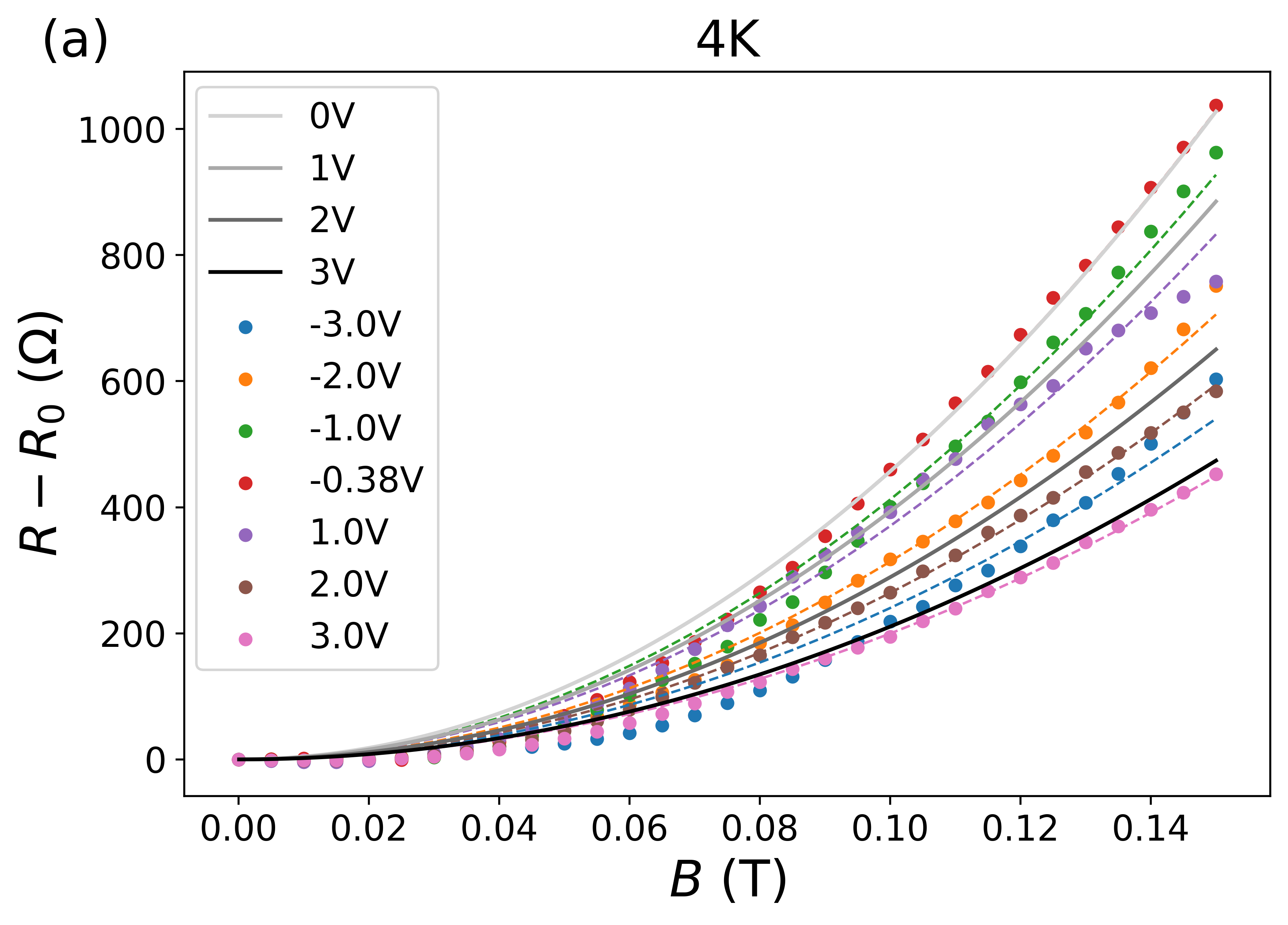}
	\includegraphics[width=.95\linewidth]{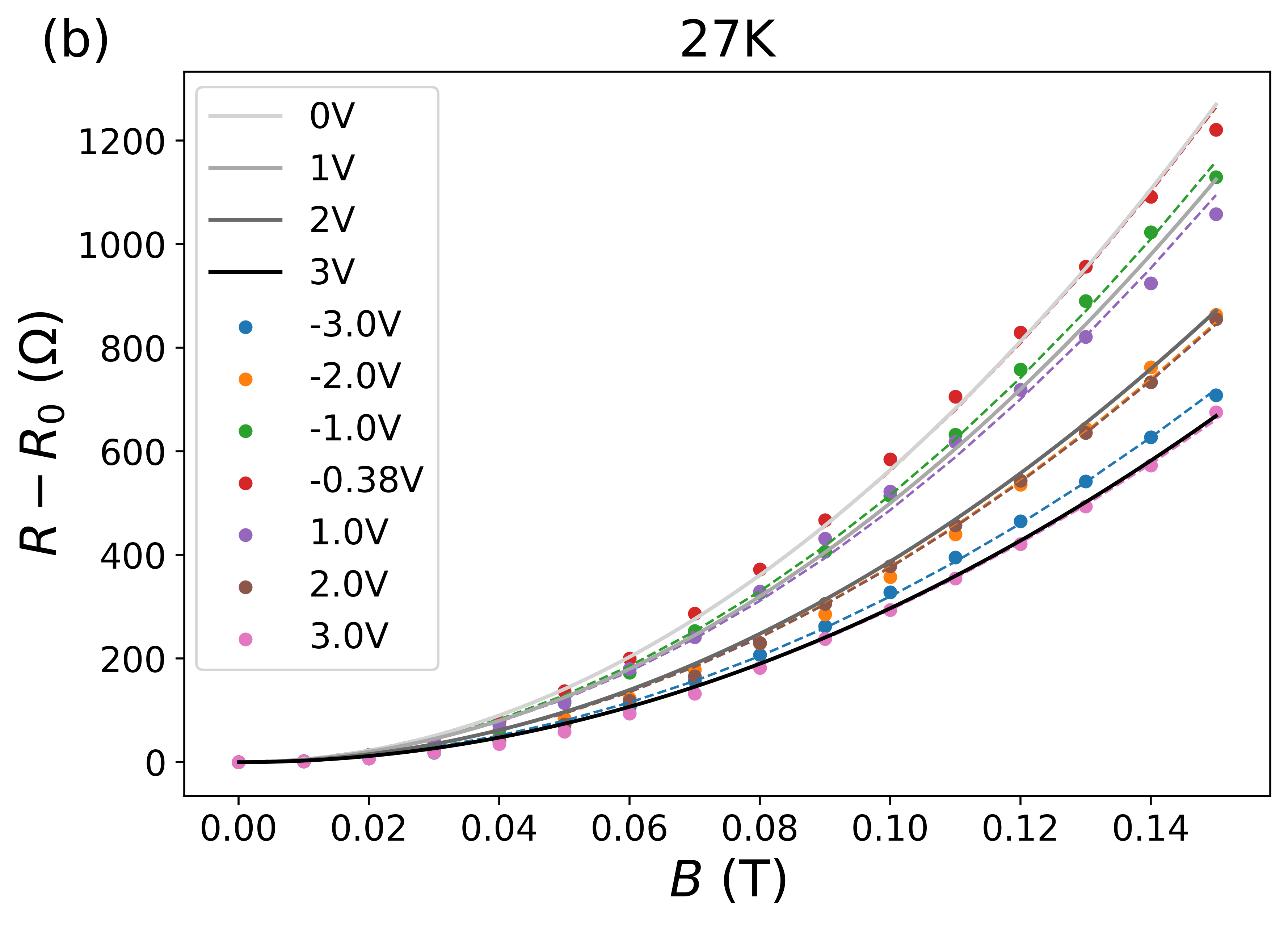}
\caption{Non-normalized magnetoresistance $R(B)-R(0)$ for 4\,K in (a) and 27\,K in (b). The points are obtained after shifting the gate voltage by $V_g^D$. The dashed lines correspond to the fitted function Eq.~(\ref{MGres}) and solid lines to theoretical zero-temperature magnetoresistance, Eq.~(\ref{eq:resistivity}), calculated using the parameters from Table \ref{tab:parameters}. Since these parameters are obtained at finite temperature and are electron-hole averaged, the dashed and solid lines do not exactly coincide for the same values of the gate voltage. \label{fig:Comparison} }
\end{figure}

Since magnetoresistance is related to mobility, the data can be employed to derive information on the impurity scattering in our sample. From the obtained fitting parameter $M$ which is given by
\begin{align}
M&=R_\text{bulk}(0)\mu_0^2=\frac{\gamma}{\pi^2 \hbar n^2}\ln\frac{r_\text{out}}{r_\text{in}}\nonumber\\
&=
\frac{1}{\pi^2 \hbar n^2}\left(\frac{1}{\gamma_s}+\frac{c(\alpha)n_\text{imp}^{c}}{n}\right)^{-1} \ln\frac{r_\text{out}}{r_\text{in}}.
\label{eq:theoryParameters}
\end{align}
In order to include the disorder-induced saturation of the density of states,  we replace $n$ with $\sqrt{n^2+n_*^2}$ in the fitting function, and then extract $\gamma_s$, $n_*$, and the effective concentration of Coulomb impurities $c(\alpha)n_\text{imp}^{C}$. The values we extract from these fits are shown in Table~\ref{tab:parameters}. 
\begin{table}
    \begin{ruledtabular}
    \begin{tabular}{ccc}
         & $T=4$\,K & $T=27$\,K  \\
         \hline
      $\gamma_s$   & 9.6 & 38.2\\
      $n_*$\ [m$^{-2}$] & 1.7$\times 10^{14}$ & 1.8$\times 10^{14}$ \\
      $c(\alpha)n_\text{imp}^C$ [m$^{-2}$] & 1.1$\times 10^{14}$ & 0.88$\times 10^{14}$
    \end{tabular}
    \caption{Parameters extracted from the fit of the data. These parameters are used in all following plots.}
    \label{tab:parameters}
    \end{ruledtabular}
\end{table}
Notably, the energy corresponding to $n_*$ is $\varepsilon_*\approx 9$\,meV, which is larger than the energies corresponding to 4\,K and 27\,K, which are 
$ 0.3$\,meV  and $ 2.3$\,meV, respectively, so that finite-$T$ corrections are small, even for $27$\,K. Moreover the value of $n_*$ is  consistent with region corresponding to disorder broadening of the Dirac point in the field effect mobility shown in Fig.~\ref{fig:FE}. In Fig.~\ref{fig:Comparison}, we show a comparison of the shifted magnetoresistance and the corresponding fit by the theoretical curves obtained from Eq.~(\ref{eq:resistivity}) with the fitting parameters from Table~\ref{tab:parameters}.

\begin{figure}[t!]
	\centering
	\includegraphics[width=.95\linewidth]{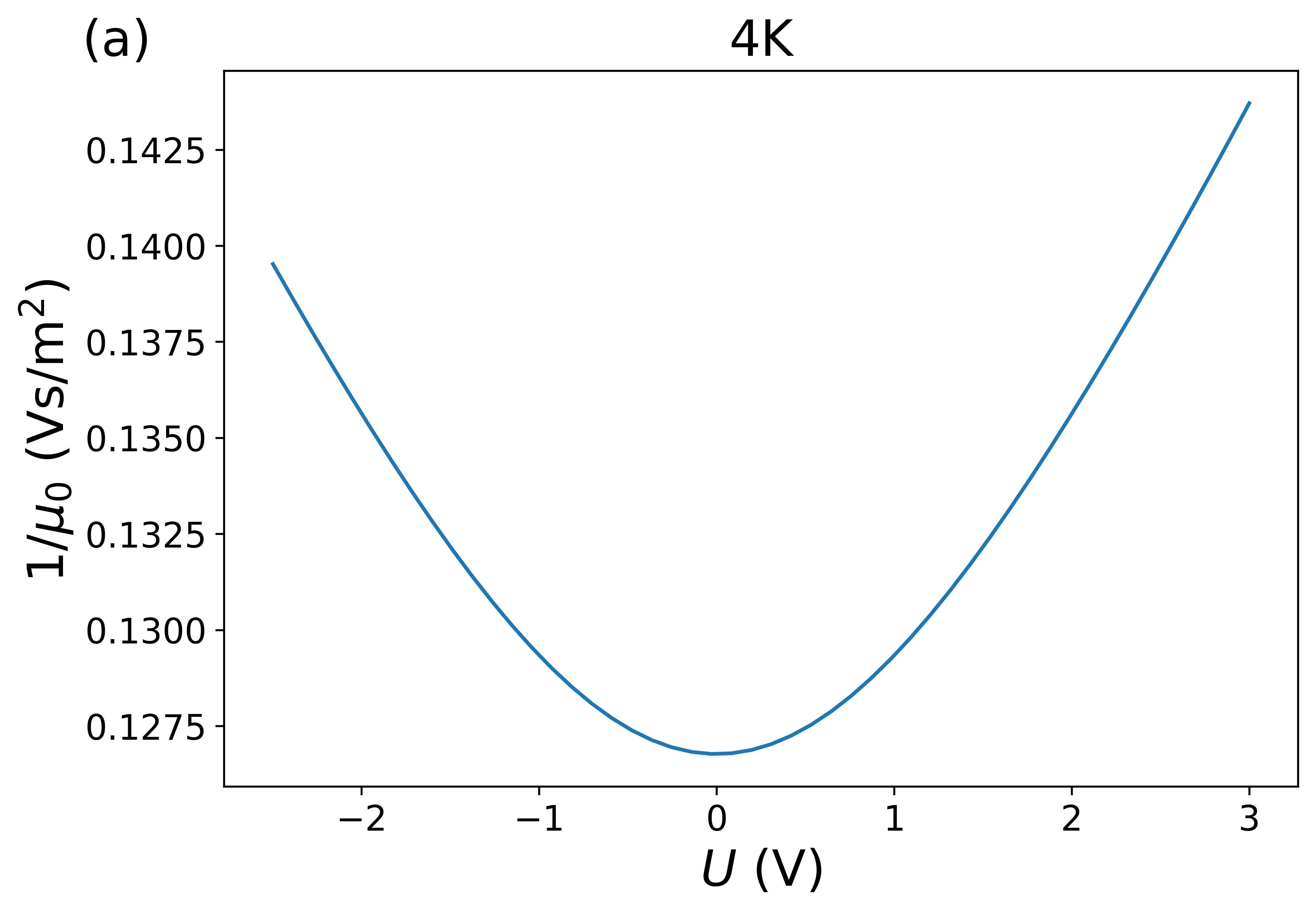}
	\includegraphics[width=.95\linewidth]{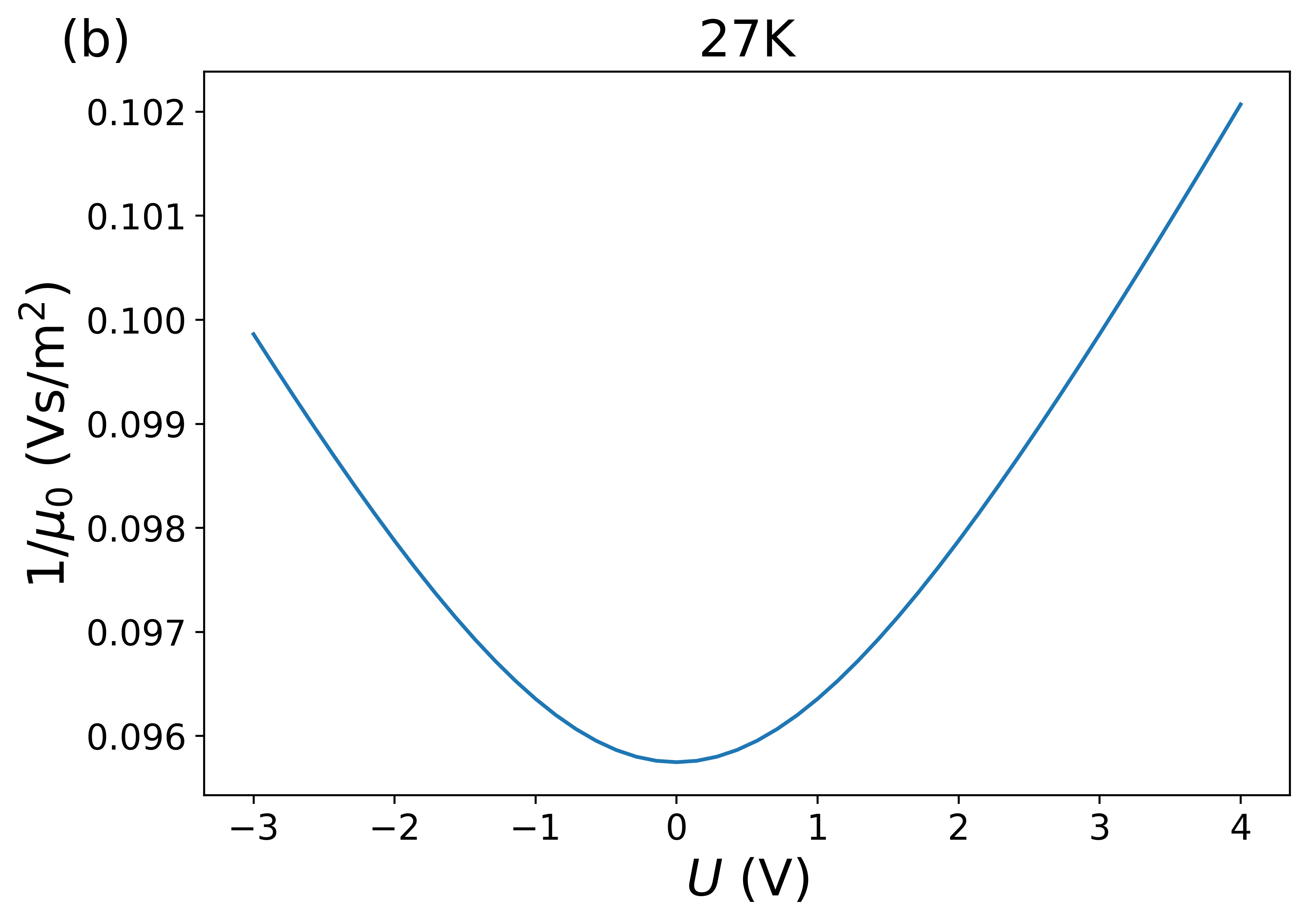}
\caption{(a) Inverse mobility, Eq.~(\ref{eq:mobility}), derived using the parameters extracted from the $4$\,K magnetoresistance data. (b) Inverse mobility derived from the $27$\,K data. The parameters are given in Table \ref{tab:parameters}. \label{fig:gammamob} }
\end{figure}

With the parameters obtained from fitting the magnetoresistance curves, we  get the mobility from Eq. (\ref{eq:mobility}), see Fig.~\ref{fig:gammamob}, where the inverse mobility $\mu_0^{-1}$ is displayed as a function of gate voltage difference $U=V_g-V_g^D$ relative to the gate voltage $V_g^D$ corresponding to the Dirac point. A clear minimum is found at the Dirac point (maximum for the mobility). The slope of the inverse mobility away from the Dirac point is determined by the strength of short-range scatterers $\gamma_s$.  We observe that at $T=27$\,K, 
the mobility varies only very slightly as a function of the gate voltage. This indicates that the role of short-range impurities is suppressed at higher temperature. Possibly, with increasing temperature residual dirt (adsorbed atoms) is removed from the sample. 

\begin{figure}[t!]
	\centering
	\includegraphics[width=.95\linewidth]{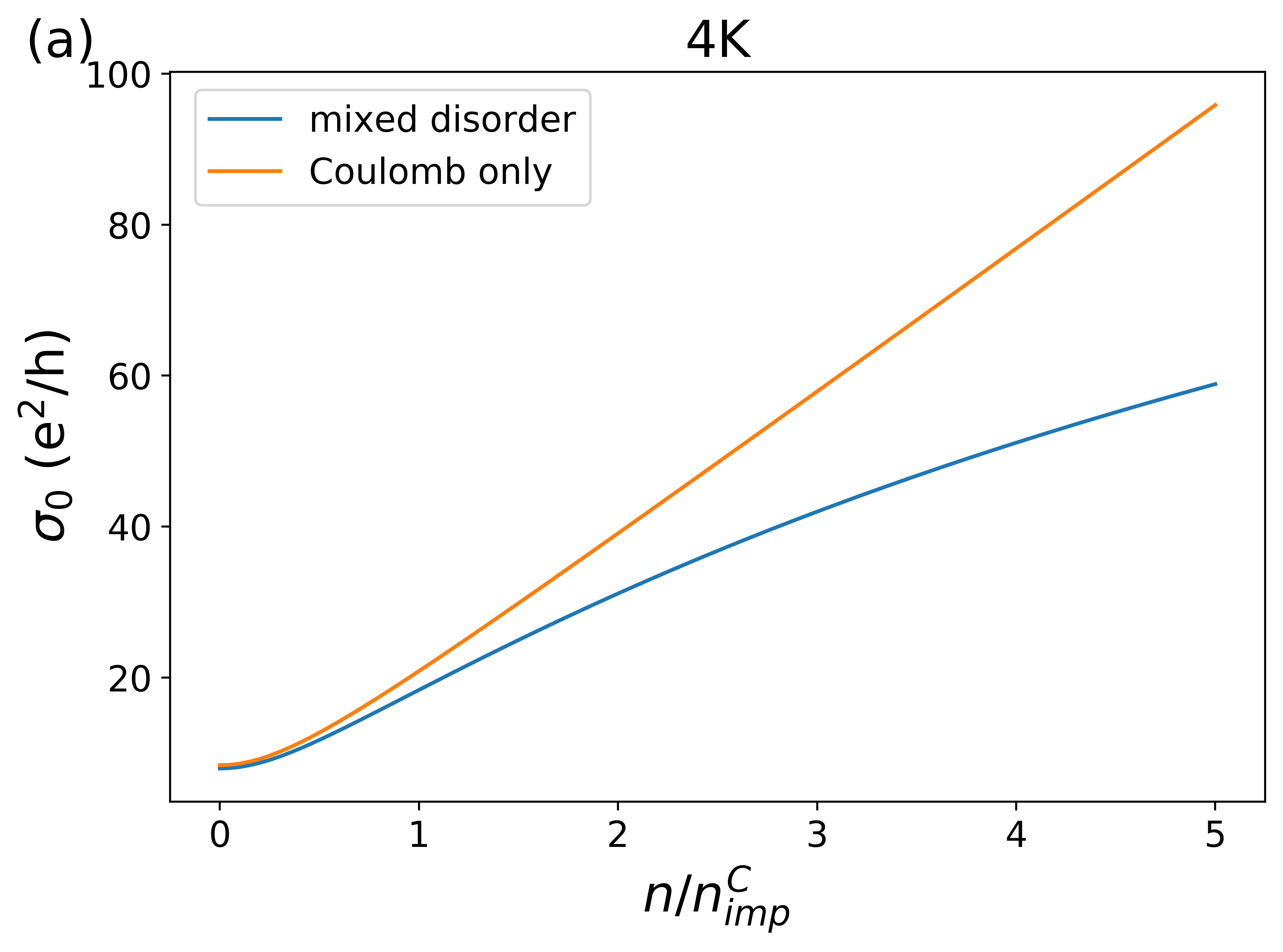}
	\includegraphics[width=.95\linewidth]{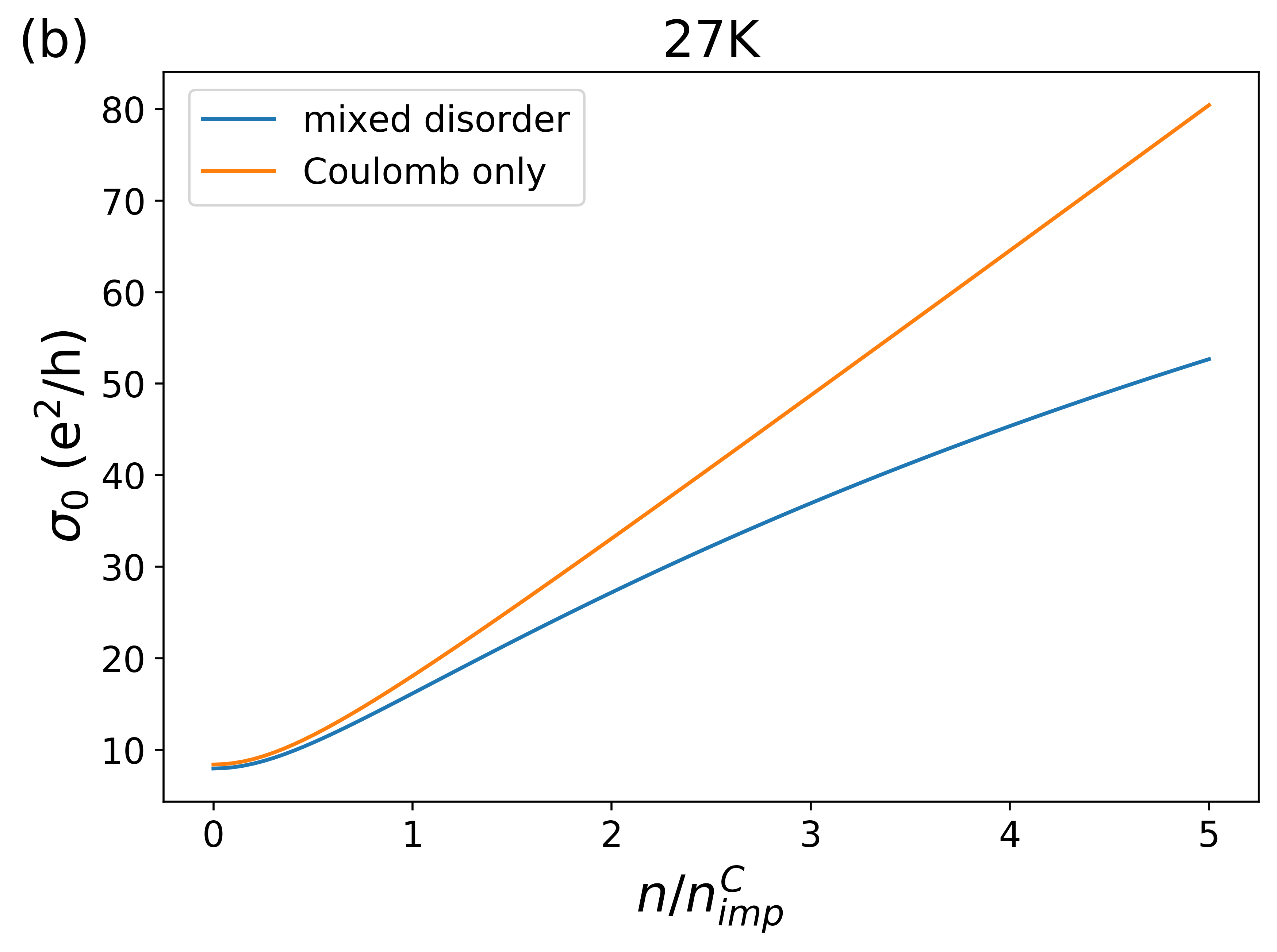}
\caption{(a) Conductivity at $B=0$T, Eq.~(\ref{eq:defGamma}), derived using the parameters extracted from the $4$\,K magnetoresistance data.  (b) Zero-field conductivity derived from the $27$\,K data. The blue curve is the result for our mixed disorder model, the orange one for $\gamma_s=\infty$, i.e., no short-range scatterers. The parameters are given in Table \ref{tab:parameters}, additionally we used $\alpha=1.3$ to determine $n_\text{imp}^C$. \label{fig:sublinear} }
\end{figure}

In order to summarize the effect of the two different types of impurities, in Fig.~\ref{fig:sublinear} we show the zero-field conductivity $\sigma_0$ determined by Eq.~(\ref{eq:defGamma}) for both our mixed disorder model and a Coulomb-impurity model. Already at experimentally accessible density ratios $n/n_\text{imp}^C\approx 2$ we observe the sublinear conductivity due to short-range scatterers discussed in Ref. \cite{Hwang2007}.

Thus, looking at the deduced impurity scattering strengths from the obtained mobilities, our magnetoresistance data should reflect effects related to both Coulomb and short-range scatterers.  Note that our Hall mobility is slightly smaller than the field-effect mobility obtained using $R(V_g)$ which yields approximately $10^5$ cm$^2$/Vs near Dirac point for the average mobility of electrons and holes. The value of mobility extracted from the analysis of the magnetoresistance is consistent with the field-effect mobility.

\begin{figure}[t!]
	\includegraphics[width=.95\linewidth]{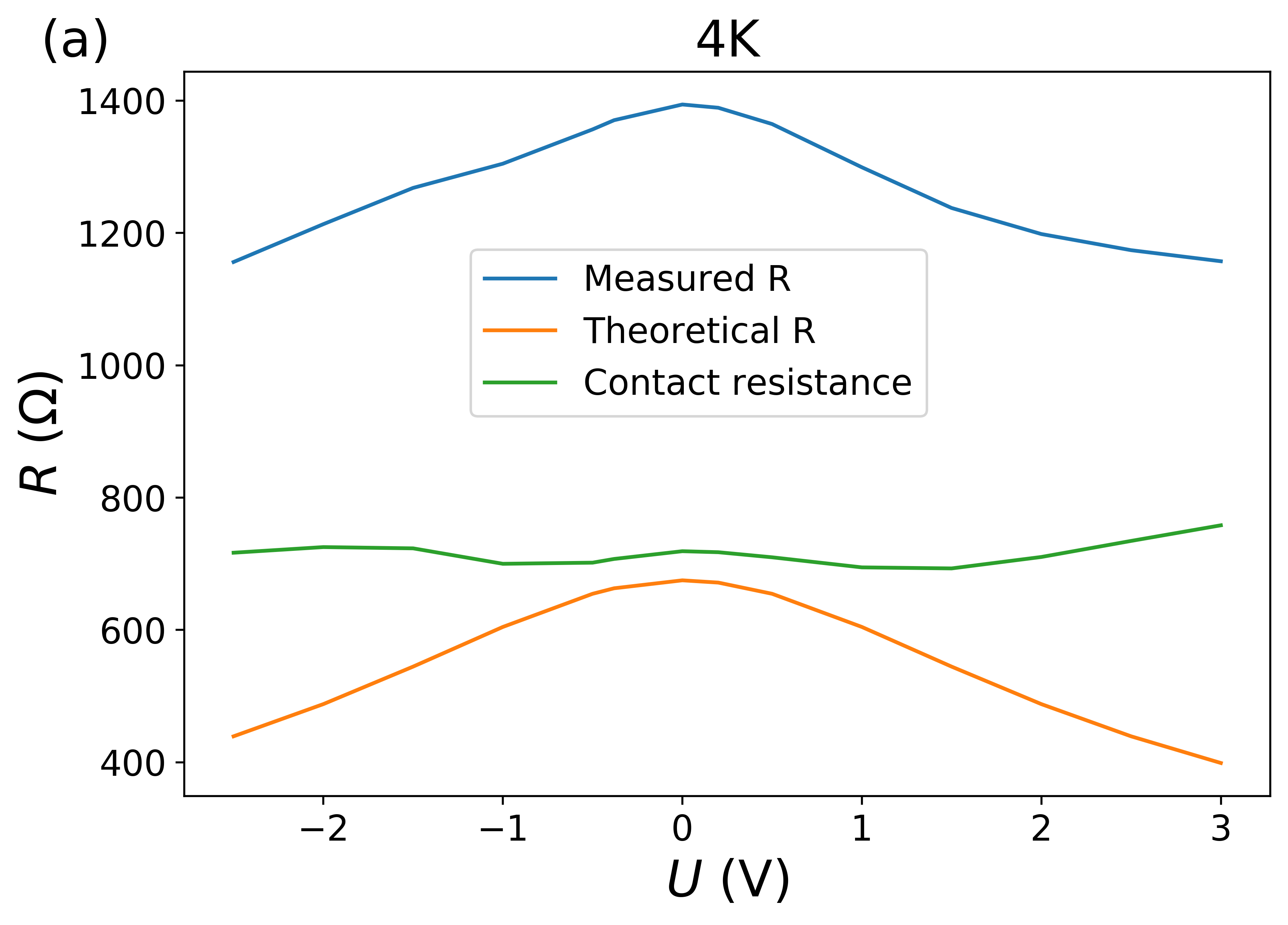}
	\includegraphics[width=.95\linewidth]{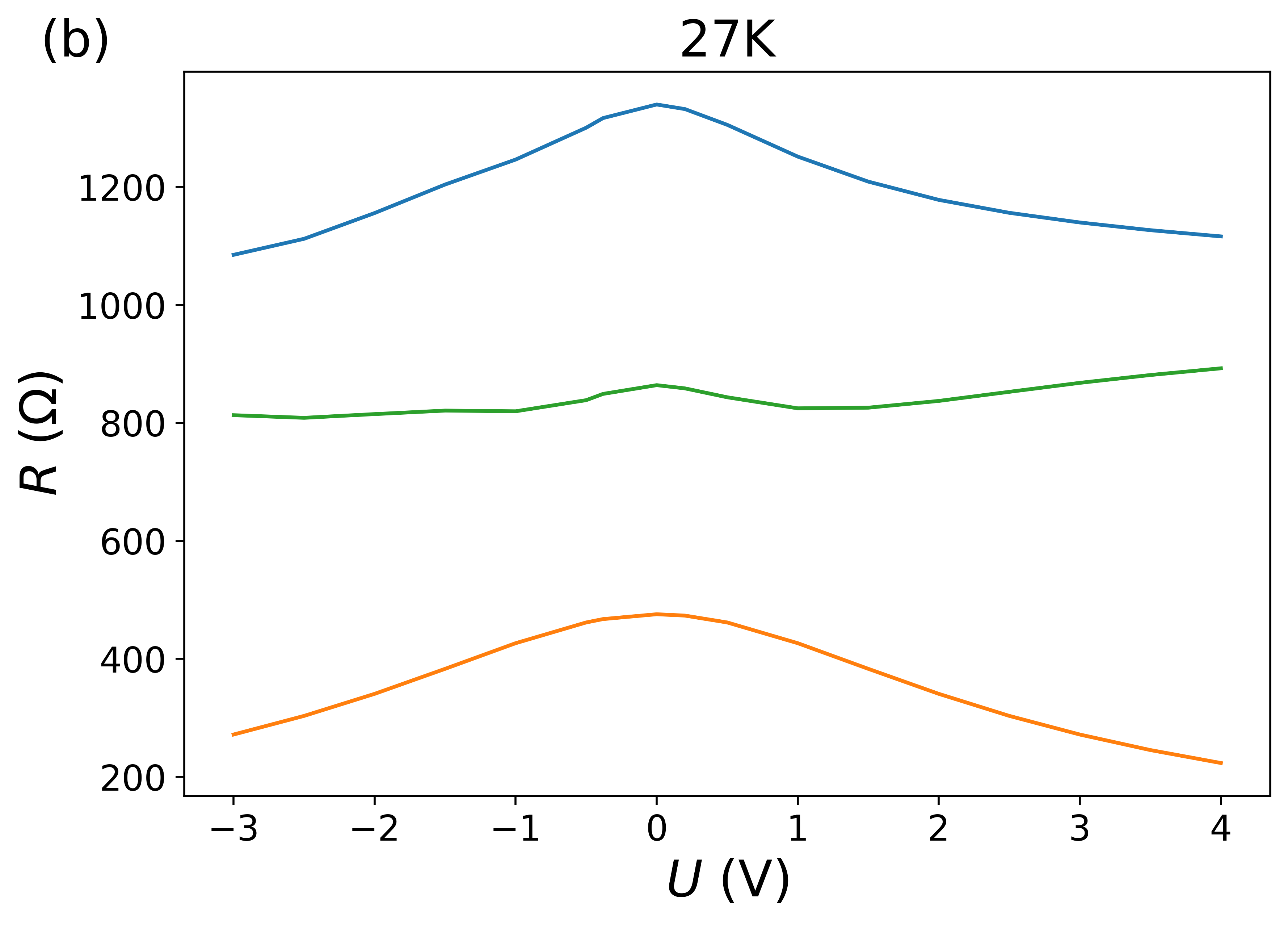}
\caption{Zero-$B$ resistance at $T=4$\,K in (a) and $T=27$\,K in (b). Blue curves: the measured resistance. Orange curves: the zero-$T$ bulk resistance calculated from  Eq.~(\ref{Corbino-resistance}) with the parameters from the fit (Table~\ref{tab:parameters}).
Green curves: the contact resistance which is obtained as the difference of the measured and bulk resistances. Note that the contact resistance is actually larger than the bulk contribution, as expected for high-mobility samples.    \label{fig:RCont} }
\end{figure}

Subtracting the disorder-induced bulk resistance with the parameters obtained from the measured magnetoresistance, we get access to the overall contact resistance, which is shown in Fig.~\ref{fig:RCont}.
The obtained value of the contact resistance is somewhat higher than the one reported for similar samples in the Corbino geometry, which can be related to the fact that the sample has been cooled down several times.

\section{Discussion and conclusions \label{sec:Discussion}}
In summary, 
we have investigated geometric magnetotoresistance in suspended graphene Corbino ring at low temperatures. The magnetoresistance,  $\Delta R(B)/R(0) \propto B^2$, is ultrastrong: it amounts to $4000 B^2$\,\%  at the Dirac point ($B$ in Tesla), with quite small temperature dependence below 30\,K. This is comparable with the ``extraordinary magnetoresistance'' in encapsulated graphene in a disk geometry observed at room temperature in Ref.~\cite{Zhou2020} (although the physical mechanism behind the low-temperature ultrastrong magnetoresistance is different). 
The relative magnetoresistance decreases with charge density and, at $|n| \simeq 3 \times 10^{-10}$ cm$^{-2}$, it is already reduced by a factor of two.

The simple zero-temperature analysis appears to be sufficient to explain the main features of the measured magnetoresistance. The quadratic magnetic field dependence is followed for both short-range and charged scatterers. 
The gate dependence of the magnetoresistance allows one to estimate the partial contributions of the short-range and long-range impurity scattering to the mobility. In particular, away from the Dirac point, the gate-voltage dependence of the mobility is entirely determined by the short-range component of the impurity potential.  At the same time, we see that transport around the neutrality point
is dominated by scattering on long-range disorder
(Coulomb impurities or ripples). However, no fundamental difference is observed between Coulomb and short-range in terms of the magnetic-field dependence of the resistance.

We find that the mobility extracted from the parabolic magnetoresistance is sufficiently high (of the order of $10^5$ cm$^2$/Vs), which is in agreement with previous estimates for similar (slightly cleaner) samples used for studying the quantum Hall effect, including the fractional quantum Hall effect. It is worth noting that, somewhat counterintuitively, the mobility is found to be higher at higher $T$.
This can be related to the technological process of preparing the sample, where the concentration of adsorbed atoms depends on temperature. The total measured resistance is given by the sum of the bulk and contact contributions. We see that in our geometry, the contact resistance is even higher than the bulk resistance, which is consistent with the high quality of the sample.

We emphasize that knowledge on the contact resistance is essential in the analysis of our geometric magnetoresistance. According to Eq. (\ref{eq:theoryParameters}), the measured magnetoresistance is proportional to $R_{\textrm{bulk}} \mu_0^2 B^2$ where $R_{\textrm{bulk}}=R-R_{\textrm{cont}}$. If we employed $R$ instead of $R_{\textrm{bulk}}$, we would overestimate the reduction of $\mu_0$ obtained from the magnetoresistance data as a function of gate voltage. For example, naively fitting the normalized data in Fig. \ref{MGresistance}b we would deduce a reduction in $\mu^2$ by a factor of two, while from Fig. \ref{fig:gammamob}b we obtain only $< 13$\% reduction in $\mu^2$. Thus, in the former case the strength of the short-range scattering would appear almost three times larger than in the correct analysis.

In the vicinity of the Dirac point at 4\,K, we find a high sensitivity of resistance variation with respect to magnetic field $dR \big/ dB=12.5$\,k$\Omega$/T at $B=0.15$\,T. According to our current noise measurements yielding $S_I=10^{-23}$\,A$^2$/Hz at 1kHz for the same sample at 10 $\mu$A, we may estimate a magnetic field sensitivity of 60 nT/$\sqrt{\textrm{Hz}}$ for our device at 4\,K. This sensitivity is excellent when compared with graphene Hall magnetometers, since our result is on par with magnetic field sensitivities of devices working in 20 times larger magnetic fields \cite{Schaefer2020}. 

To conclude, the analysis of graphene magnetoresistance at different values of the gate voltage in the Corbino geometry allowed us to extract information about scattering mechanisms in the sample, and to separate the bulk and contact contributions to the resistance. The strong magnetoresistance of Corbino geometry at low magnetic fields appears to be a very powerful tool of characterization of the graphene samples.
As an outlook, it will be very interesting to investigate experimentally and theoretically the magnetoresistance of graphene in this geometry at elevated temperatures, when electronic hydrodynamic effect become pronounced.

\section*{Acknowledgements}
We thank Pavel Alekseev, Dmitry Golubev, and Sheng-Shiuan Yeh for fruitful discussions and comments. This work was supported by the Academy of Finland projects 314448 (BOLOSE), 310086 (LTnoise) and 312295 (CoE, Quantum Technology Finland), by ERC (grant no. 670743), as well as by the DFG within FLAG-ERA Joint Transnational Call (Project GRANSPORT) and RFBR (grant no. 20-02-00490).  This research project utilized the Aalto University OtaNano/LTL infrastructure which is part of European Microkelvin Platform EMP (funded by European Union’s Horizon 2020 Research and Innovation Programme Grant No. 824109).

\appendix

\section{Dirac point shift\label{sec:Diracshift}}

In the resistance vs. gate voltage data, Fig.~\ref{fig:bothRes}, we observe a shift of the 
resistance maximum (associated with the position of the Dirac point) with increasing magnetic field.
We attribute this shift to the dependence of the screening length on the magnetic field, which is only relevant for small densities. 
In order to be able to fit the quadratic magnetoresistance, one first has to get rid of this shift of the Dirac point.

\begin{figure}[t!]
	\centering
	\includegraphics[width=.95\linewidth]{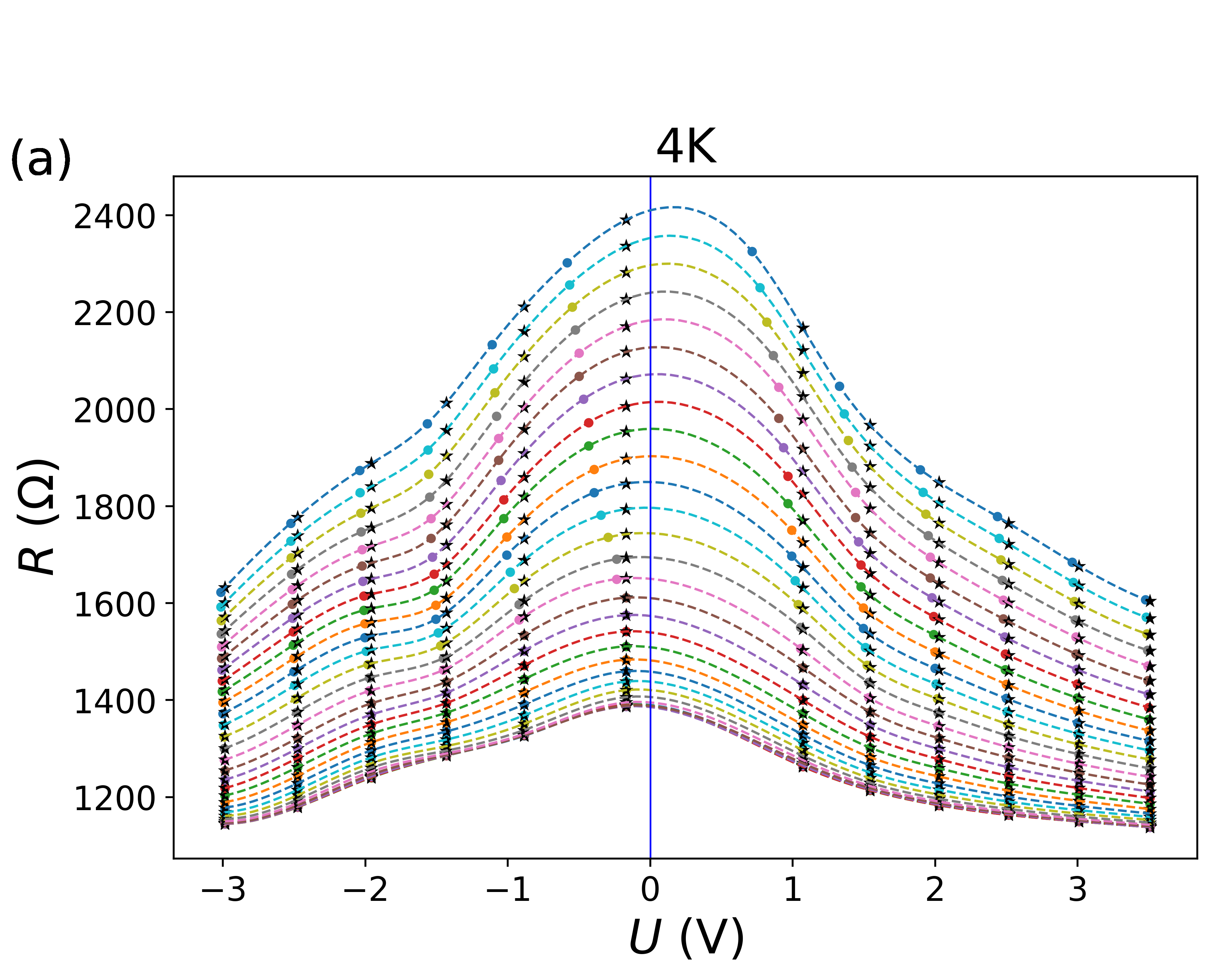}
	\includegraphics[width=.95\linewidth]{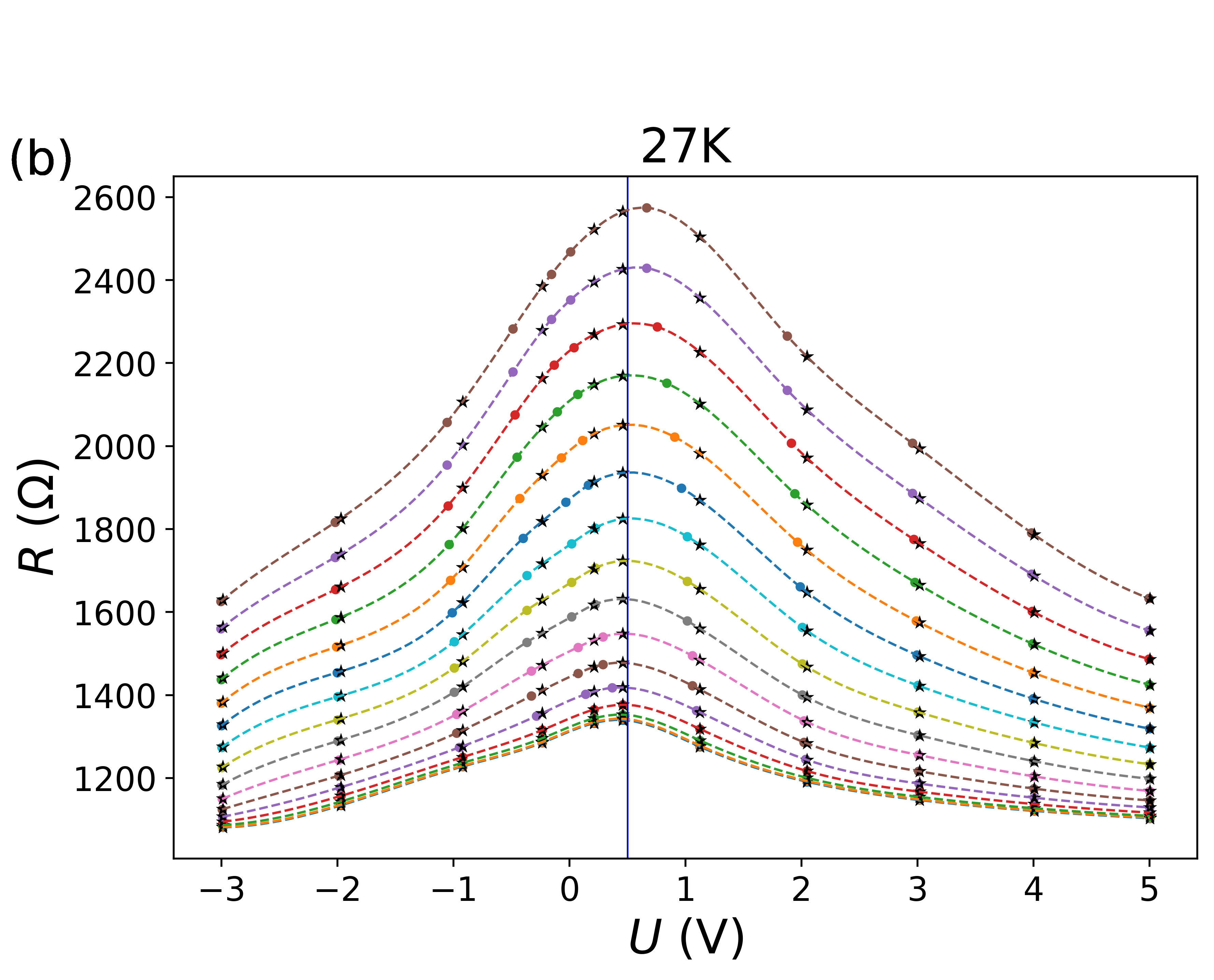}
\caption{Resistance at $4$\,K (a) and $27$\,K (b), shifted according to Eq.~(\ref{eq:shift}). Colored dots are the shifted measured points and black stars show the data obtained from the spline. (a): $V_2=1$V and $V_1=0$; (b): $V_2=1$V and $V_1=0.5$V. \label{fig:ResShifted} }
\end{figure}

Within the picture based on the effect of magnetic field on screening, one should only shift data points close to the Dirac point, while not affecting those farther away from it. Since we have data only for a discrete set of values of the gate voltage,
in order to determine the gate voltage corresponding to the Dirac point, we find the maximum of the $R(V)$ curve given by a cubic spline interpolation of the measured data points. For each value of magnetic field, this maximum $V_0$ is shifted to the same voltage $V_1$. For all other data points, we adopt a phenomenological Ansatz, where, away from the Dirac point, the shift reduces exponentially, i.e., a measured point on the curve corresponding to $B$ is shifted by
\begin{align}
V\Rightarrow V-(V_0-V_1)e^{-\frac{|V-V_0|}{V_2}}. \label{eq:shift}
\end{align}
The voltage $V_1$ is the same for all curves and is determined by the maximum of the most symmetric curve in the unshifted case, and $V_2$ describes the characteristic voltage window where the effect of magnetic field on screening is seen. Since the shift for different magnetic fields will generically be different, one also has to use the spline interpolation to get access to the resistance at the same voltages for all magnetic fields. We take this voltages to be the ones on the shifted zero magnetic field curve. A result for thus shifted and unshifted data points is shown in Fig.~\ref{fig:ResShifted}.

Comparing Fig.~\ref{fig:ResShifted} with Fig.~\ref{fig:bothRes} of the main text, we see that
the shifted curves are indeed more symmetric, which supports the idea of stronger shift around the neutrality point as a result of magnetic-field effect on screening.
We note, however, that a homogeneous (independent of the distance to the Dirac point) shift, used for the actual fitting procedure, captures correctly the shift near the neutrality point. Since this is the main manifestation of the screening-induced shift in magnetoresistance curves (the difference in the tails of magnetoresistance is not that important for the fitting procedure), we adopt the simplest shifting in our analysis. Note that, in all cases, the small feature close to $-2$V (attributed to a quasi-resonance in scattering by adsorbed local impurities) is preserved after shifting the curves.

\section{Charged impurities \label{sec:AppCharge}}

In the main text, we gave the transport and quantum scattering times for short-range impurities, Eq.~(\ref{tau0}), and Coulomb impurities, Eq.~(\ref{tau1}). Here we show how to derive these expressions for large energies. The behaviour close to the neutrality point is discussed in Sec.~\ref{sec:AppSat}.

Very generally, the scattering rates can be written as 
\begin{align}
\frac{1}{\tau_{i}(\varepsilon)}&=\frac{n_\text{imp}\nu(\varepsilon)}{8\hbar} \int_0^{2\pi}\mathrm{d}\theta\,
\left|V\left(q(\theta)\right)\right|^2\nonumber\\
&\times\begin{cases}
\sin^2\theta, & i=\text{tr}\\
(1+\cos\theta), & i=q
\end{cases}  \label{eq:generalScattering},
\end{align}
where $\nu(\varepsilon)$ is the density of states and we express the transferred momentum as $$q(\theta)=2(\varepsilon/\hbar v)|\sin(\theta/2)|$$ with $\theta$ being the scattering angle. For short-range impurities, $V(q)=U_0$ and $n_\text{imp}=n_\text{imp}^s$; the integration over $\theta$ then leads exactly to Eq.~(\ref{tau0}). In the case of charged impurities, we consider the screened interaction potential in the random-phase approximation:
\begin{align}
V(q) =\frac{2\pi e^2/q \epsilon_\infty}{1+(2\pi e^2/q\epsilon_\infty) \Pi(q,0)}.
\end{align}
Here $\epsilon_\infty$ is the background dielectric constant and $\Pi(q)$ is the static polarization operator which is given by the thermodynamic density of states ($\upmu$ is the chemical potential):
\begin{align}
\lim_{q\rightarrow 0}\Pi(q)=\frac{\partial n}{\partial \upmu}.
\end{align}
This quantity can be connected to the single-particle density of states via
\begin{align*}
n(\upmu)&=\int \mathrm{d}\varepsilon\, n_F(\varepsilon)\nu(\varepsilon) \nonumber\\
\Rightarrow \ \frac{\partial n}{\partial \upmu}
&= \int \mathrm{d}\varepsilon\left(-\frac{\partial n_F(\varepsilon)}{\partial \varepsilon}\right) \nu(\varepsilon).
\end{align*}
As mentioned in the main text,the density of states can in general be deduced from Eq.~(36) in Ref.~\cite{Alekseev2013}. From there we see that 
away from the neutrality point, the oscillatory magnetic-field dependent corrections to the compressibility are exponentially suppressed and hence not seen in the experiment. 
Moreover, since we are at temperatures below than the chemical potential away from the neutrality point, we can use a zero $T$ approximation and get 
\begin{align}
\lim_{q\rightarrow 0}\Pi(q,0)\approx\frac{2\upmu}{\pi v_F^2\hbar^2}=\nu(\upmu).
\end{align}
By introducing the effective interaction strength $$\alpha = \frac{e^2}{\hbar v_F \epsilon_\infty},$$ 
we can bring the screened Coulomb interaction to the form
\begin{align}
V(q)&=\frac{2\pi e^2}{\epsilon_{\infty}(q+K)},\\
K &= \frac{2\pi e^2 N}{\epsilon_{\infty}}\lim_{q\rightarrow 0}\Pi(q,0)\approx 2 \alpha \pi\hbar v_F \nu(\upmu). \label{eq:inverseScreeningLength}
\end{align}
We get thus get for the scattering rates:
\begin{align}
\frac{1}{\tau_{i}^{C}(\varepsilon)}&=
\frac{\pi^2}{4}\hbar^3v_F^4\alpha^2n_\text{imp}^{C}\nu(\varepsilon)\nonumber\\
&\times\int_0^{\pi}\frac{\mathrm{d}\theta}{\left(\varepsilon\sin\frac{\theta}{2}+\alpha\nu(\upmu)\pi\hbar^2 v_F^2\right)^2 }\nonumber\\
&\times\begin{cases}
\sin^2\theta, & i=\text{tr}\\
1+\cos\theta, & i=q
\end{cases}.
\end{align}
At low temperatures, the typical energies are very close to the chemical potential and we can set $\varepsilon\to \upmu$ in the transferred momentum
$q$ in the interaction matrix element:
\begin{align}
\frac{1}{\tau_\text{tr}^{C}(\varepsilon)}&\approx\frac{\pi}{4}\alpha^2 v_F^2\hbar n_\text{imp}^C\varepsilon\int_0^{2\pi}
\frac{\mathrm{d}\theta\, \sin^2\theta}{[\upmu|\sin(\theta/2)|+\alpha N \upmu/2]^2}\nonumber\\
&=\frac{\pi}{2}v^2\hbar n_\text{imp}^C\frac{\varepsilon}{\upmu^2} c(\alpha),\\
c(\alpha)&=\alpha^2 \int_0^{\pi}\frac{\mathrm{d}\theta \sin^2\theta}{(\sin(\theta/2)+\alpha N /2)^2}.
\end{align}
This can be brought into the form
\begin{align}
\tau_\text{tr}^{C}(\varepsilon)=\frac{2\gamma^{C} \hbar}{\varepsilon},
\quad \gamma^{C} = \frac{\upmu^2}{\pi v_F^2\hbar^2 n_\text{imp} c(\alpha)},
\end{align}
which is exactly Eq.~(\ref{tau1}) for zero temperature, where $\upmu=\varepsilon_F$. Using relation (\ref{eq:density}), this is also exactly $\gamma_C$ as introduced in Eq.~(\ref{eq:gamma_C}).
Similarly, we get for the quantum scattering rate

\begin{align}
\frac{1}{\tau_{q}^{C}}&\approx\frac{\pi}{2}\alpha^2 v_F^2\hbar n_\text{imp}^C
\frac{\varepsilon}{\upmu^2}
\int_0^{\pi}\frac{\mathrm{d}\theta\,(1+\cos\theta)}
{[\sin(\theta/2)+\alpha N /2]^2}\nonumber\\
&=\frac{\pi}{2}v_F^2\hbar n_\text{imp}^C\frac{\varepsilon}{\upmu^2} d(\alpha),\\
d(\alpha)&=\alpha^2 \int_0^{\pi}\frac{\mathrm{d}\theta (1+\cos\theta)}{(\sin(\theta/2)+\alpha N /2)^2}
\end{align}
which again is of the form
\begin{align}
\tau_q^{C}(\varepsilon)=
\frac{\gamma_C^{\prime}\hbar}{\varepsilon},
\quad \gamma_C^{\prime} = \frac{2 \upmu^2}{\pi v_F^2\hbar^2 n_\text{imp}^C d(\alpha)},
\end{align}
as used in Eq.~(\ref{tau1}).
The values of $c(\alpha)$ and $d(\alpha)$ for some realistic $\alpha$ are given by
\begin{align}
c(\alpha)&=\begin{cases}
0.14, & \alpha = 0.5\\
0.22, & \alpha = 1.0\\
0.26, & \alpha = 1.5\\
0.29, & \alpha = 2
\end{cases}\ , \nonumber\\
d(\alpha)&=\begin{cases}
0.43, & \alpha = 0.5\\
0.55, & \alpha = 1.0\\
0.61, & \alpha = 1.5\\
0.65, & \alpha = 2
\end{cases}\ . \label{eq:somevalues}
\end{align}
Thus, for intermediate $\alpha$, the relation between the two scattering rates, given by  $\frac{d(\alpha)}{2}\approx c(\alpha)$, is similar to that for short-range scattering. 
Whenever we use a specific value of $\alpha$, we chose $\alpha=1.3$, thus accounting for the renormalization of velocity, as well as for the 
screening by the metallic parts of the setup. 


In a general setup, both short-range and Coulomb scatterers are present. Here we discuss how this mixture affects the scattering times for energies away from the Dirac point, where the density of states is not affected by disorder. We will stick with the assumption of no inter-valley scattering, diagonality in the sublattice space, and no correlations between different kinds of scattering. 
This corresponds to summing up the self-energies, where we would not consider mixed diagrams, and translates into a sum rule for transport times, Eq.~(\ref{eq:totaltransporttime}).
We will further assume that in the relevant limit, the density of states is not modified by the magnetic field, since corrections are exponentially suppressed. Then we can use the transport times as written in Eq.~(\ref{tau0}) and Eq.~(\ref{tau1}) to write the total transport time $\tau_\text{tr}$ as
\begin{align}
\tau_\text{tr}(\varepsilon)=\frac{2\gamma \hbar}{|\varepsilon|},
\quad \gamma=\frac{\gamma_C \gamma_s}{\gamma_C+\gamma_s}. 
\end{align}
The conductance kernel, as found from the Boltzmann equation, is then given by
\begin{align}
\sigma_{xx}(\varepsilon) &= \frac{e^2 v_F^2}{2}\frac{\tau_\text{tr}(\varepsilon) \nu(\varepsilon) }
{1+\left[\omega_c(\varepsilon)\tau_\text{tr}(\varepsilon)\right]^2 } 
= \frac{\sigma_0}{1+[\omega_c(\varepsilon)\tau_{\text{tr}}(\varepsilon)]^2}\nonumber\\
&=\frac{2 e^2\gamma}{\pi\hbar}\frac{1}{1+\left(\frac{2\gamma \hbar^2\Omega^2}{\varepsilon^2}\right)^2}\label{eq:CondKernel},
\end{align}
where $\Omega=v_F/\ell_B$, leading to Eq.~(\ref{eq:sigmaxxzeroT}) of the main text.

\section{Finite-temperature effects  \label{sec:AppT}}

In the main text, we restricted ourselves to the zero temperature limit of Eq.~(\ref{eq:TDependence}). Here we discuss finite-temperature corrections in the Drude formula, as well as effects due to electron-electron interaction (EEI) which also introduce finite-temperature corrections to Eq.~(\ref{eq:sigmaxxzeroT}) and thus Eq.~(\ref{eq:resistivity}). 
Let us consider, for the illustrative purpose,
the case of only short-range potential.

The problem generally has two energy scales, the first one is the temperature $k_B T$, and the second one is the energy $\varepsilon_m$ introduced by the product $\omega_c(\varepsilon)\tau_\text{tr}(\varepsilon)$ which we write as
\begin{align}
\omega_c(\varepsilon)\tau_\text{tr}(\varepsilon) = \frac{2eB\hbar v_F^2 }{\varepsilon^2} \equiv\frac{\varepsilon_m^2}{\varepsilon^2}
\end{align}
We will restrict ourselves to low temperatures 
$k_B T \ll \upmu$ and small magnetic fields $,\varepsilon_m\ll \upmu$
For finite but low temperatures and small fields, the integral of the conductance kernel Eq.~(\ref{sigma_xx}) over energies 
can be brought into the form \cite{Alekseev2013}:
\begin{align}
\sigma_{xx}&=\sigma_0\int_{-\infty}^{\infty}\mathrm{d}\varepsilon\left(-\frac{\partial n_F(\varepsilon)}{\partial\varepsilon}\right) \left(1-\frac{\varepsilon_m^4}{\varepsilon^4+\varepsilon_m^4}\right)\nonumber\\
&=\sigma_0\left[1-\int_{-\infty}^{\infty}\mathrm{d}\varepsilon\left(-\frac{\partial n_F(\varepsilon)}{\partial\varepsilon}\right) \frac{\varepsilon_m^4}{\varepsilon^4+\varepsilon_m^4}\right],
\end{align}
we used Eq.~(\ref{tau0}) and Eq.~(\ref{tau1}) for the transport scattering time. 

One notices that both the derivative of the Fermi function and the fraction in the second expression are peaked. The derivative of the Fermi function is peaked around $\varepsilon=\upmu$ with a width given by the temperature, while the term $\varepsilon_m^4/(\varepsilon^4+\varepsilon_m^4)$ is peaked around $\varepsilon=0$ and its width is determined by $\varepsilon_m$. 
Since we assume that both $T,\varepsilon_m \ll \upmu$, these peaks are well separated, and the integral can be written as a sum of the contributions of the two peaks, i.e.,
$
\sigma_{xx}=\sigma_0-\left(\sigma_{xx}^{(T)}+\sigma_{xx}^{(\varepsilon_m)}\right). 
$
%

For low temperatures, the derivative of the Fermi function has a finite width of the order of $T$ around $\varepsilon=\upmu$. To incorporate finite-temperature corrections to the conductivity, we thus expand the fraction in powers of $(\varepsilon-\upmu)$ 
and after evaluating the integral with get
\begin{align}
\sigma_{xx}^{(T)}&\approx\sigma_0
\left[\frac{\varepsilon_{m}^{4}}{\upmu^4+ {\varepsilon_m} ^4}-\frac{2\pi^2 T^2 \upmu^2 {\varepsilon_m}^4}{3}
\frac{\left(3 {\varepsilon_m}^4-5\upmu ^4 \right)}
{\left(\upmu^4+{\varepsilon_m} ^4\right)^3}\right]. \label{eq:condexact}
\end{align}
The first term of this expression is the only non-vanishing contribution at zero $T$ and leads exactly to Eq.~(\ref{eq:sigmaxxzeroT}).


The contribution of the second peak is found by fixing the value of the Fermi function at its value at  $\varepsilon=0$ and then evaluating the integral:
\begin{align}
\sigma_{xx}^{(\varepsilon_m)}\simeq \sigma_0\varepsilon_m\frac{e^{-\upmu/T}}{\sqrt{2}T}.
\end{align}
Thus the total Drude resistivity for short-range impurities  is given, in the regime of low temperatures and low magnetic fields, by
%
%
%
\begin{align}
&\rho_{xx}=\frac{1}{\sigma_{xx}}\approx\frac{1}{\sigma_0}\\
&\times\left[1+
\frac{{\varepsilon_m}^4}{\upmu^4} \left(1+\frac{10 \pi ^2 T^2}{3 \upmu^2}\right)-
\frac{{\varepsilon_m}^8}{\upmu^8}\frac{16 \pi ^2  T^2}{3 \upmu^{2}}+\frac{\pi  {\varepsilon_m}}
{\sqrt{2}T}  e^{-\frac{\upmu}{T}} \right]\nonumber.
\end{align}
From this expressions we see, that the mobility $\mu_0$ itself does not acquire finite temperature corrections, since they all require finite magnetic field. However the $B$ dependence of the magnetoresistance does acquire additional terms (in particular, the $B^4$ term which is absent at $T=0$,
as well as the $\sqrt{B}$ term \cite{Alekseev2013} which is, however, exponentially suppressed at low $T\ll \upmu$). Furthermore, the $B^2$ dependence is also slightly modified by a finite temperature.
Finite-$T$ corrections would have a similar structure if one includes Coulomb impurities.

\subsection{Effects of electron-electron interaction \label{sec:AppEEI}}

The effect of electron-electron interaction (EEI) on the magnetoresistance of graphene was explored in Refs. \cite{Kozikov2010, Jobst2012}. 
Since EEI does not influence $\sigma_{xy}$, we can directly  employ their main result, which is the EEI correction to magnetoresistivity:
\begin{align}
\Delta \rho_\text{EEI}&=\left[ (\omega_c\tau_\text{tr})^2-1\right] \frac{e^2\rho_0^2}{2\pi^2 \hbar}A\, \ln\frac{k_B T \tau_{\text{tr}}}{\hbar}, \label{eq:EEICorrection}
\end{align}
where
\begin{align}
A &= 1+c\left[1-\frac{\ln(1+F_0^{\sigma})}{F_0^{\sigma}}\right],
\\
F_0^{\sigma} &= -\alpha\int_0^{2\pi} \frac{\mathrm{d}\theta}{2\pi} \frac{\cos^2\theta/2}{\sin\theta/2+2\alpha},
\end{align}
and $c$ is the number of multiplets. Depending on the temperature, $c=3$, 7, or 15 for very low, moderately low, and high temperatures, respectively. It describes the number of ungapped non-singlet two-particle states contributing to Hartree-type correction to the conductivity. Since each electron has a well defined spin and valley quantum number, which can both take two values, there are in total 16 possibilities for two particle states, one of which will always be a spin and valley singlet. In fact, since there may be inter- or intra-valley scattering, valley is not necessarily a good quantum number in this sense, depending of the hierarchy of the temperature, the  intra-valley phase breaking time $\tau_*$, and the inter-valley scattering time $\tau_\text{iv}$. For $k_B T<\hbar/\tau_*$, channels mixing different spins do not contribute, thus there remain $2\times 4=8$ channels, of which one is a singlet, i.e., $c=7$. If $k_B T<\hbar/\tau_\text{iv}$, valley is not a good quantum number anymore, thus we get $4$ states, of which one is a multiplet, i.e., $c=3$. 
According to Ref. \cite{Jobst2012}, the relation $\tau_*<\tau_\text{iv}$ is usually fulfilled. 
Below we give the numerical values of the Fermi liquid constant $F_0^{\sigma}$ for some values of $\alpha$:
\begin{align}
    F_0^{\sigma} = \begin{cases}
    -0.18, & \alpha = 0.5\\
    -0.21, & \alpha = 1\\
    -0.22, & \alpha = 1.5\\
    -0.23, & \alpha = 2
    \end{cases}\label{eq:fermiLiqConst}\ .
\end{align}

\begin{figure}[t!]
	\centering
	\includegraphics[width=.95\linewidth]{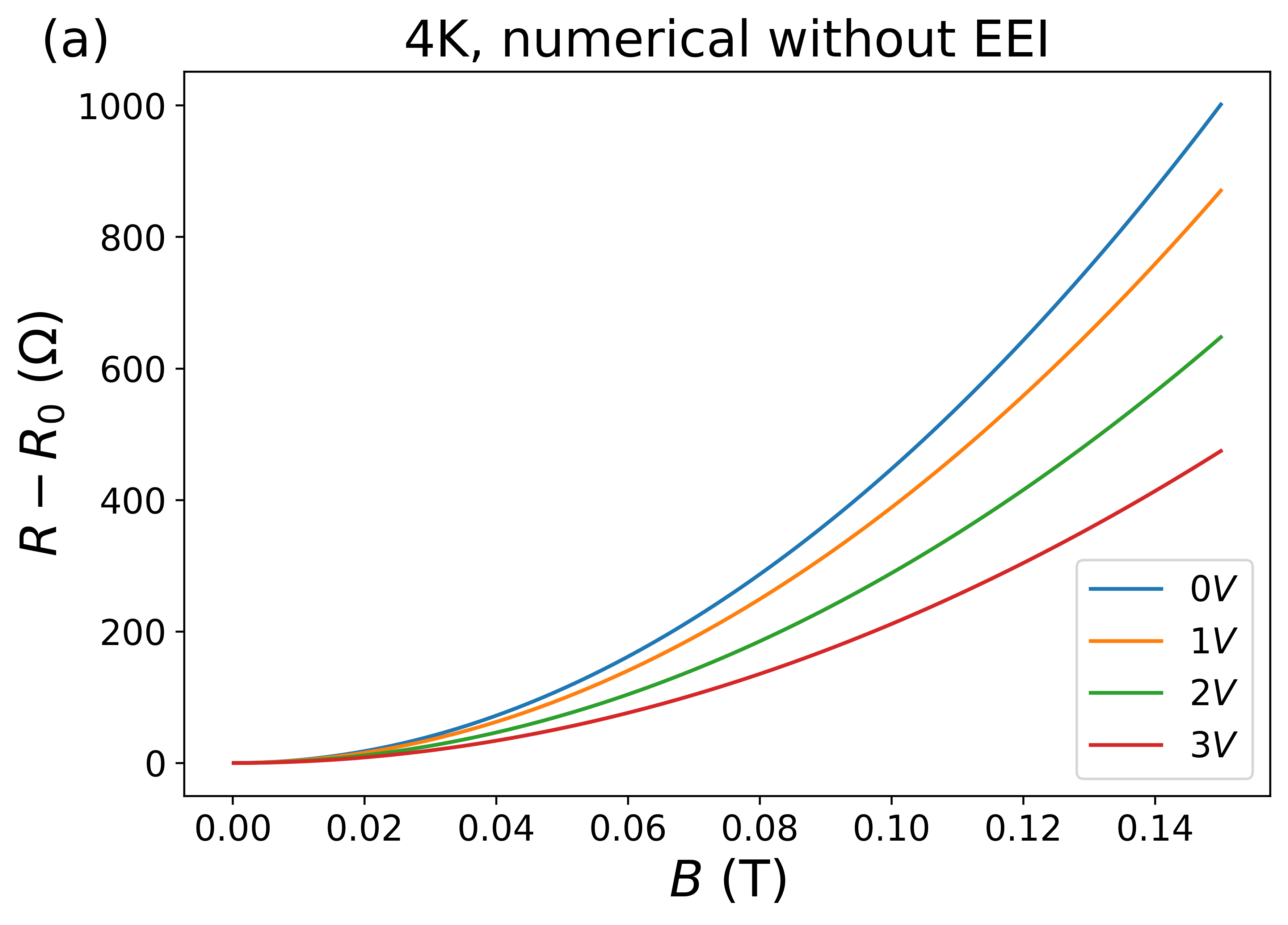}
	\includegraphics[width=.95\linewidth]{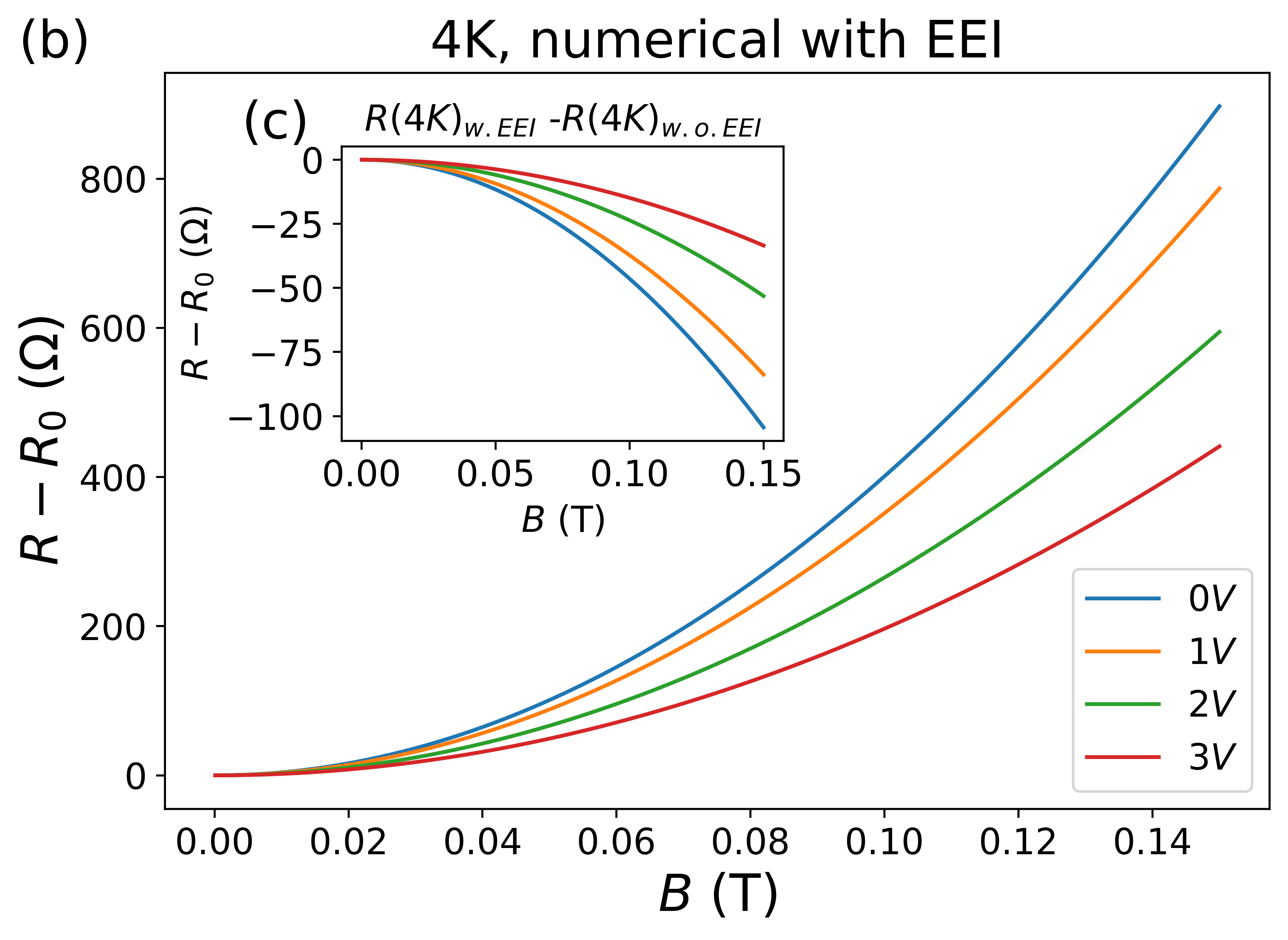}
\caption{Magnetoresistance at $T=4$\,K with parameters from the fit (Table \ref{tab:parameters}) without the EEI correction (a) and including it (b). The EEI correction leads to a small suppression of the magnetoresistance, without changing the functional form. This negative correction is shown in inset (c). Here $\alpha=1.3$ and $c=3$ were chosen.  \label{fig:EEIeffects} }
\end{figure}

\begin{figure}[t!]
	\centering
	\includegraphics[width=.95\linewidth]{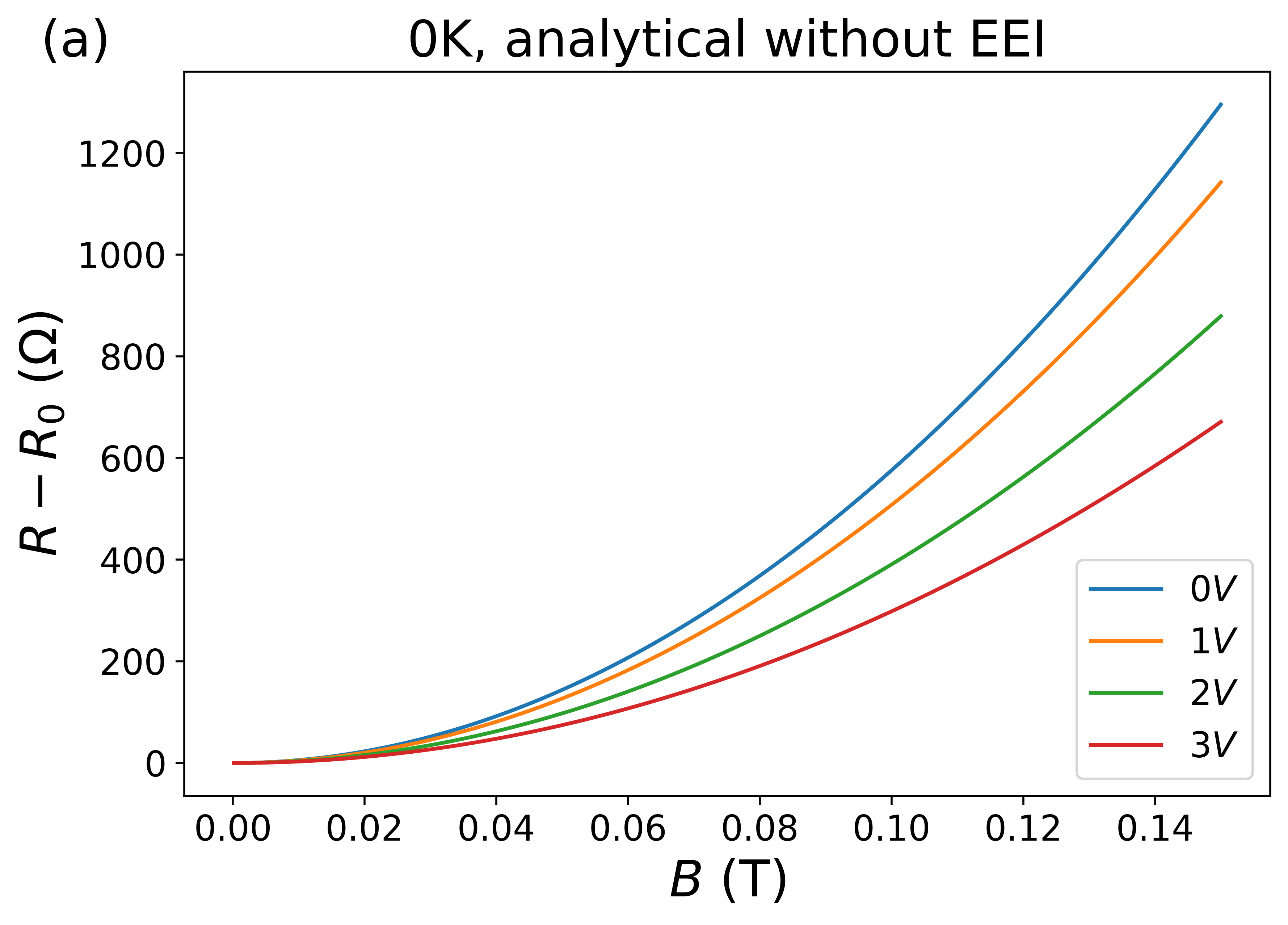}
	\includegraphics[width=.95\linewidth]{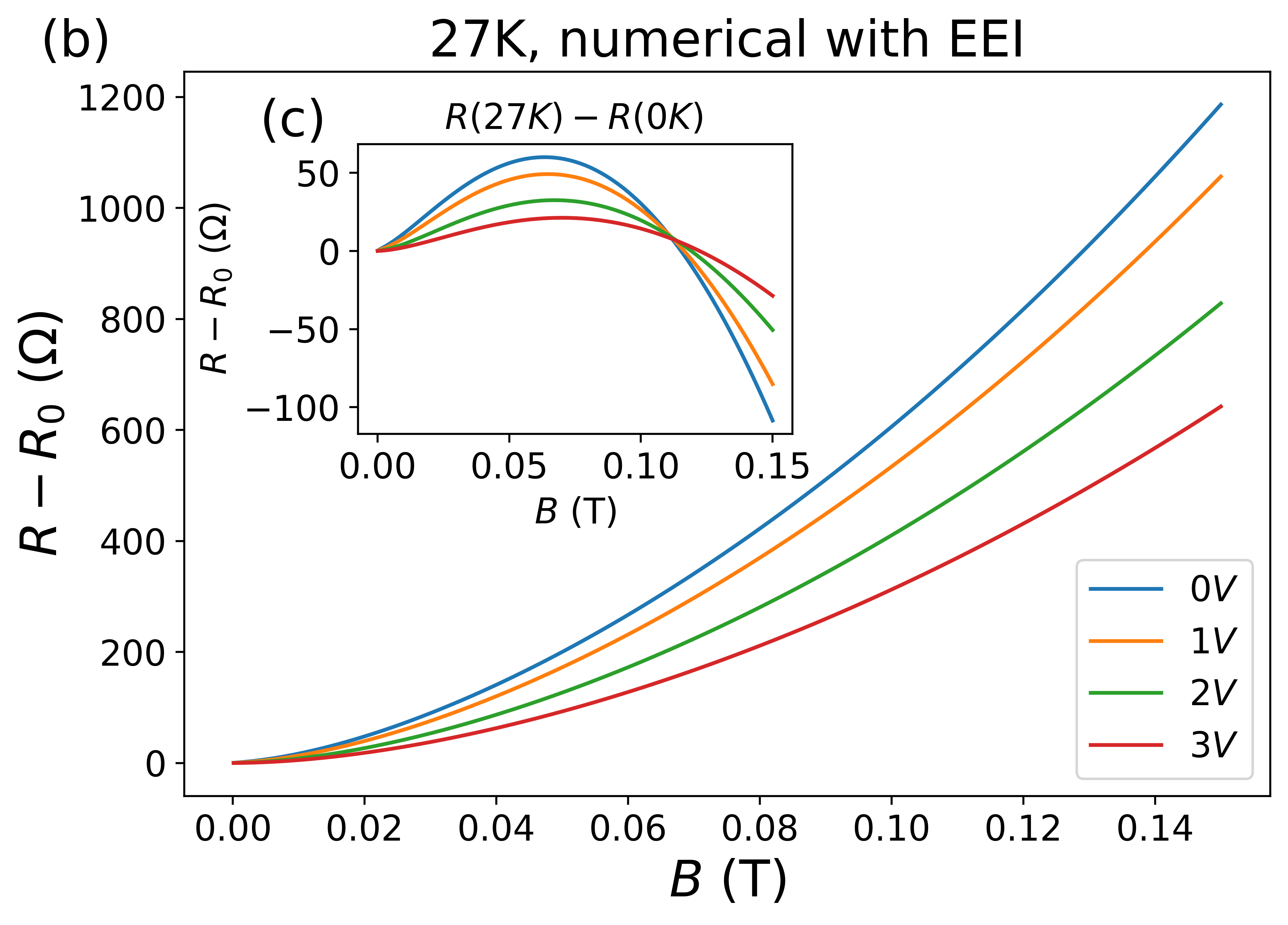}
\caption{(a) Magnetoresistance obtained from the zero-temperature solution (\ref{eq:resistivity}) with the parameters from the fit (Table \ref{tab:parameters}) for $T=27$\,K. (b) Magnetoresistance obtained by using the parameters from the fit and solving Eq.~(\ref{eq:TDependence}) numerically, with the EEI correction included according to Eq. (\ref{eq:EEICorrection}) ($\alpha=1.3$ and $c=7$). The inset (c) shows the finite temperature corrections to the magnetoresistance. Depending on the magnitude of the magnetic field this correction is either positive or negative and it also changes the functional form very slightly. \label{fig:MB} }
\end{figure}

Correction (\ref{eq:EEICorrection}) does influence the mobility and it also influences the prefactor of the $B^2$ dependence. For low temperatures, this is a negative correction.  The effect of EEI alone on the resistance at $T=4$\,K is shown in Fig.~\ref{fig:EEIeffects}. 
By plugging the value of $\gamma$ obtained from the fit of magnetoresistance in the analytical expression for zero $T$, Eq. (\ref{eq:resistivity}), we obtain the left panel of Fig.~\ref{fig:MB}.
For comparison, in the right panel of Fig.~\ref{fig:MB}, we show the result of numerical evaluation of Eq.~(\ref{eq:TDependence}) at $T=27$\,K,
with the EEI correction included. 
A comparison  with  the  zero-$T$ plot  in Fig.~\ref{fig:MB} shows that the EEI correction at $T=27$\,K is almost negligible.  At the same time, the EEI  produces  the leading finite-$T$ correction at $T=4$\,K. Overall, we observe that the effect of finite temperature is rather weak in the considered range of parameters.
In combination, finite temperature does change the mobility and it also influences the magnetic field behaviour of the resistance, so that formula (\ref{eq:resistivity}) is no longer exact and the relation (\ref{eq:theoryParameters}) does not, strictly speaking, yield the true parameters. However, as we see from Figs.~\ref{fig:EEIeffects} and \ref{fig:MB}, these effects are rather small in the experimentally accessed range, which justifies the neglect of these effects.

\section{Vicinity of the Dirac point \label{sec:AppSat}}

As seen in Fig.~\ref{fig:bothRes}, the resistance does not diverge at the Dirac point, as would be expected from combining  the resistivity $\rho_0=1/\sigma_0$  from Eq.~(\ref{eq:defGamma}) with the effective $\gamma$ from Appendix \ref{sec:AppCharge} using $\gamma_C$ as expressed in Eq.~(\ref{eq:gamma_C}). The reason for this is the saturation of the density of states close to the Dirac point due to disorder \cite{CastroNeto2009,Ostrovsky2006}.  Below a certain chemical potential $\upmu_*$, the quasiparticle pole in the Green's function is effectively absent, and all quantities should be fixed below this value. The relevant scale for this behaviour is given by Eq.~(\ref{eq:saturation}). In order to determine the changes this induces and find Eq.~(\ref{eq:nstar}), we discuss here how to find the relevant scale $\mu_*$ from the condition
\begin{align}
\frac{\hbar}{2\tau_q(\upmu_*)} =\upmu_*. \label{eq:conditionequal}
\end{align}
Below $\upmu_*$, the density of states saturates, while it is not affected for larger energies, i.e.
\begin{align}
\nu(\varepsilon)&=\begin{cases}
\nu_*, & |\varepsilon|\ll\upmu_*,\\
\nu_0(\varepsilon), & |\varepsilon|\gg\upmu_*
\end{cases}.
\end{align}
This value enters directly into the calculation of all scattering rates Eq.~(\ref{eq:generalScattering}); it also directly determines the screening radius in the Coulomb impurity case Eq.~(\ref{eq:inverseScreeningLength}). While it is clear, that Eq.~(\ref{eq:generalScattering}) is only true if the density of states is not strongly broadened, the idea is to approach the crossover from the side of large energies, where this is the case.  We can then calculate the Coulomb scattering rates as follows:
\begin{align}
\frac{1}{\tau_\text{tr}^{C}(\upmu)}
&=\frac{\pi^2}{4}\hbar^3v_F^4n_\text{imp}^{C}c(\alpha)\begin{cases}
\dfrac{\nu_*}{\upmu_*^2}, & \upmu\ll\upmu_*\\[0.2cm]
\dfrac{\nu_0(\upmu)}{\upmu^2}, & \upmu\gg\upmu_*
\end{cases}\ ,
\\
\frac{1}{\tau_{q}^{C}(\upmu)}
&=\frac{\pi^2}{4}\hbar^3v_F^4n_\text{imp}^{C}d(\alpha)\begin{cases}
\dfrac{\nu_*}{\upmu_*^2}, & \upmu\ll\upmu_*\\[0.2cm]
\dfrac{\nu_0(\upmu)}{\upmu^2}, & \upmu\gg\upmu_*
\end{cases}
\ ,
\end{align}
and, in a similar fashion, we get for short-range scatterers: 
\begin{align}
\frac{1}{\tau_{\text{tr}}^{s}(\upmu)}
&=\frac{1}{\gamma_s}\frac{\pi \hbar v_F^2}{4}\begin{cases}
\nu_*, & \upmu\ll\upmu_*\\
\nu_0(\upmu), & \upmu\gg\upmu_*
\end{cases}\ ,\\
\frac{1}{\tau_{q}^{s}(\upmu)}
&=\frac{1}{\gamma_s}\frac{\pi \hbar v_F^2}{2}\begin{cases}
\nu_*, & \upmu\ll\upmu_*\\
\nu_0(\upmu), & \upmu\gg\upmu_*
\end{cases}\ .
\end{align}
By assuming that the expression for  $\upmu\gg\upmu_*$ is still reasonably close for $\upmu\sim \upmu_*$  and using both short-range and Coulomb impurities in Eq.~(\ref{eq:conditionequal}) we find
\begin{align}
\upmu_* &=\sqrt{\frac{\pi}{2}v^2\hbar^2 d(\alpha)\frac{n_\text{imp}^{C}}
{2-1/\gamma_s}},\nonumber\\
n_* 
&= \frac{d(\alpha)n_\text{imp}^{C}}
{4-{2/\gamma_s}}.\label{eq:nstarexact}
\end{align}
which leads to the approximation Eq.~(\ref{eq:nstar}). This value depends on the density of charged impurities and $\gamma_s$. 
The effective $\gamma$ can then be extracted from the definition of $\gamma$, Eq.~(\ref{eq:defGamma}). We see that the density of states actually drops out, only the influence of $\nu(\varepsilon)$ on the electronic density (via $\upmu^2$) is relevant. We find the required asymptotics 
\begin{align}
\frac{1}{\gamma}=\frac{1}{\gamma_s}+\begin{cases}
\dfrac{ c(\alpha)n_\text{imp}^{C}}{n_*}, & n\ll n_*\\[0.2cm]
\dfrac{c(\alpha) n_\text{imp}^{C}}{n}, & n\gg n_*
\end{cases},
\end{align}
which we phenomenologically fulfill by tweaking the relation  between density and chemical potential in graphene to 
\begin{align}
\sqrt{n^2+n_*^2} 
= N\frac{\upmu^2}{4\pi\hbar^2v^2},
\label{eq:disorderBroadening}
\end{align}
and consequently we replace $n$ by $\sqrt{n^2+n_*^2}$ in all fits and plots, as mentioned in the main text.

\begin{figure}[h!]
	\centering
	\includegraphics[width=.95\linewidth]{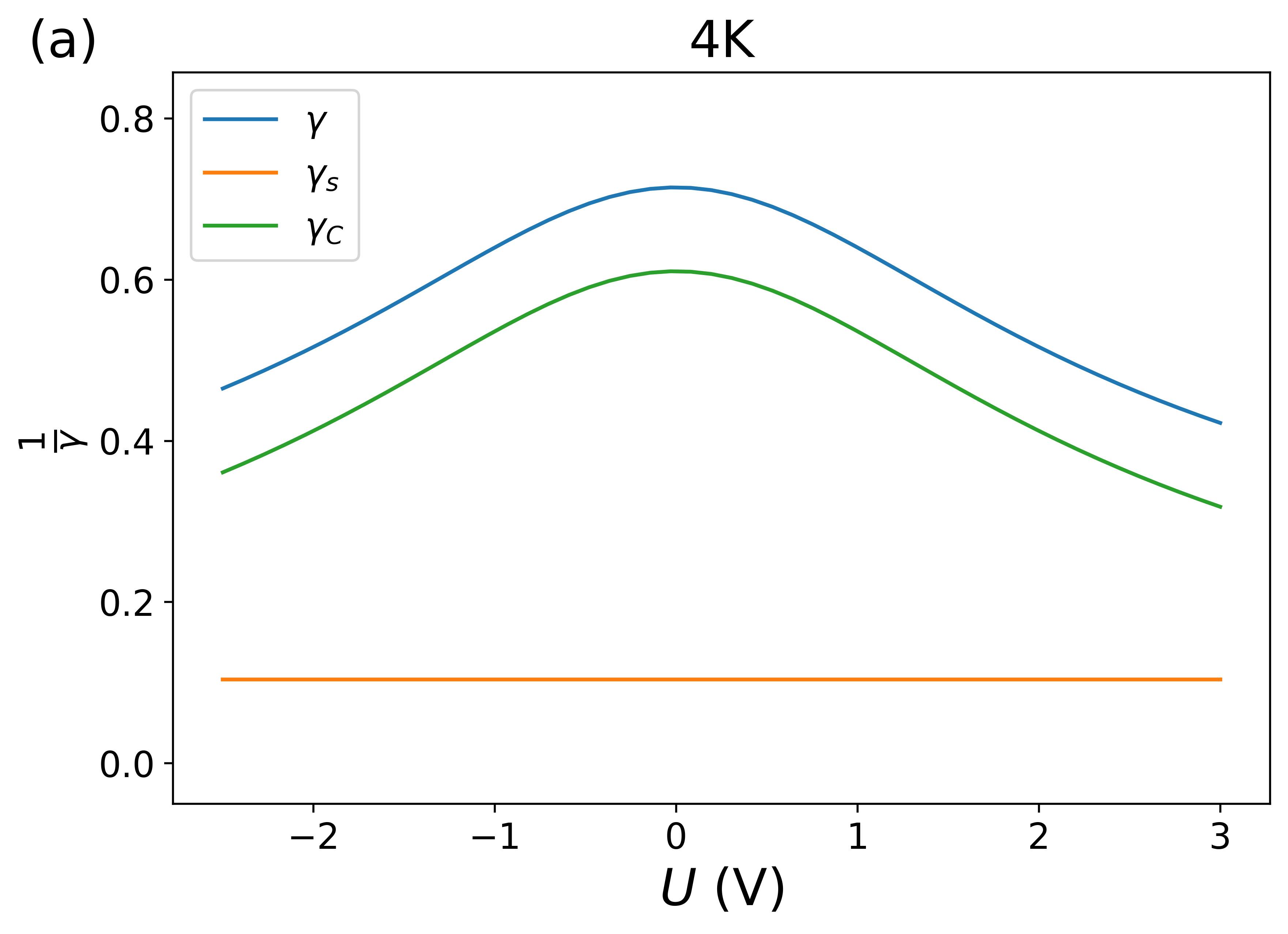}
	\includegraphics[width=.95\linewidth]{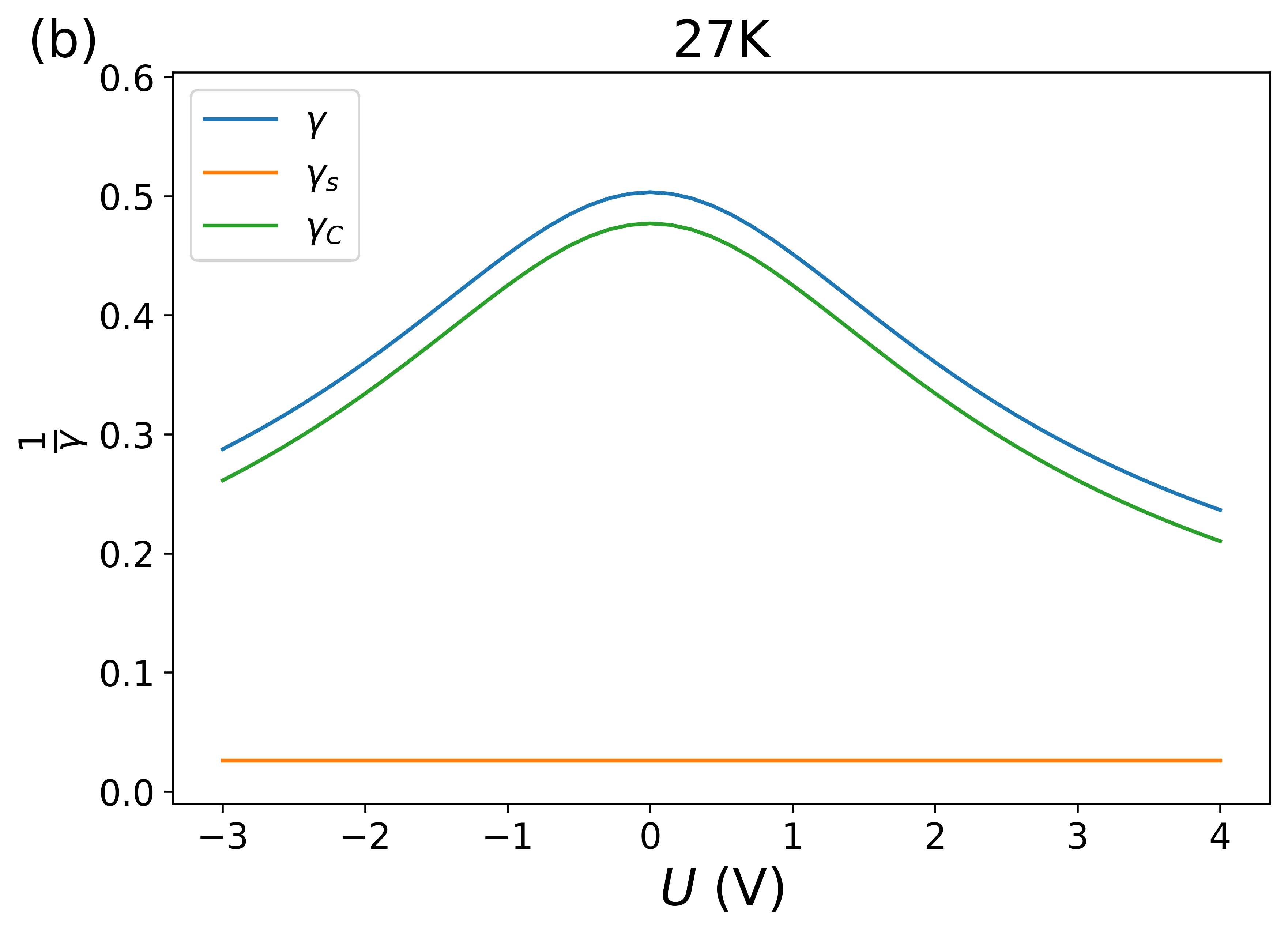}
\caption{Effective disorder parameter $1/\gamma$ defined by Eq. (\ref{eq:defEffGamma}) and its two parts $1/\gamma_C$ and $1/\gamma_s$ for $T=4$\,K (a) and $T=27$\,K (b). The parameter $\gamma$ is smaller for 4\,K, which means that disorder is stronger at lower temperatures. The parameters are given in Table \ref{tab:parameters}. 
\label{fig:gammas} }
\end{figure}

Using the general conductivity formula  (\ref{eq:sigmaxxzeroT}) with the finite disorder broadening, Eq.~(\ref{eq:disorderBroadening}), we get the broadened form of Eq.~(\ref{eq:mobility}) and Eq.~(\ref{eq:gamma_C}):
\begin{align}
\frac{1}{\mu_0}&\approx \frac{\pi \hbar}{2e}\left(\frac{\sqrt{n^2+n_*^2}}{\gamma_s}+c(\alpha)n_\text{imp}^{C}\right)
\end{align}
and 
\begin{align}
\frac{1}{\gamma} 
= \frac{1}{\gamma_s}+\frac{c(\alpha)n_\text{imp}^{C}}{\sqrt{n^2+n_*^2}}\label{eq:defEffGamma}.
\end{align}

\begin{figure}[t!]
	\includegraphics[width=.95\linewidth]{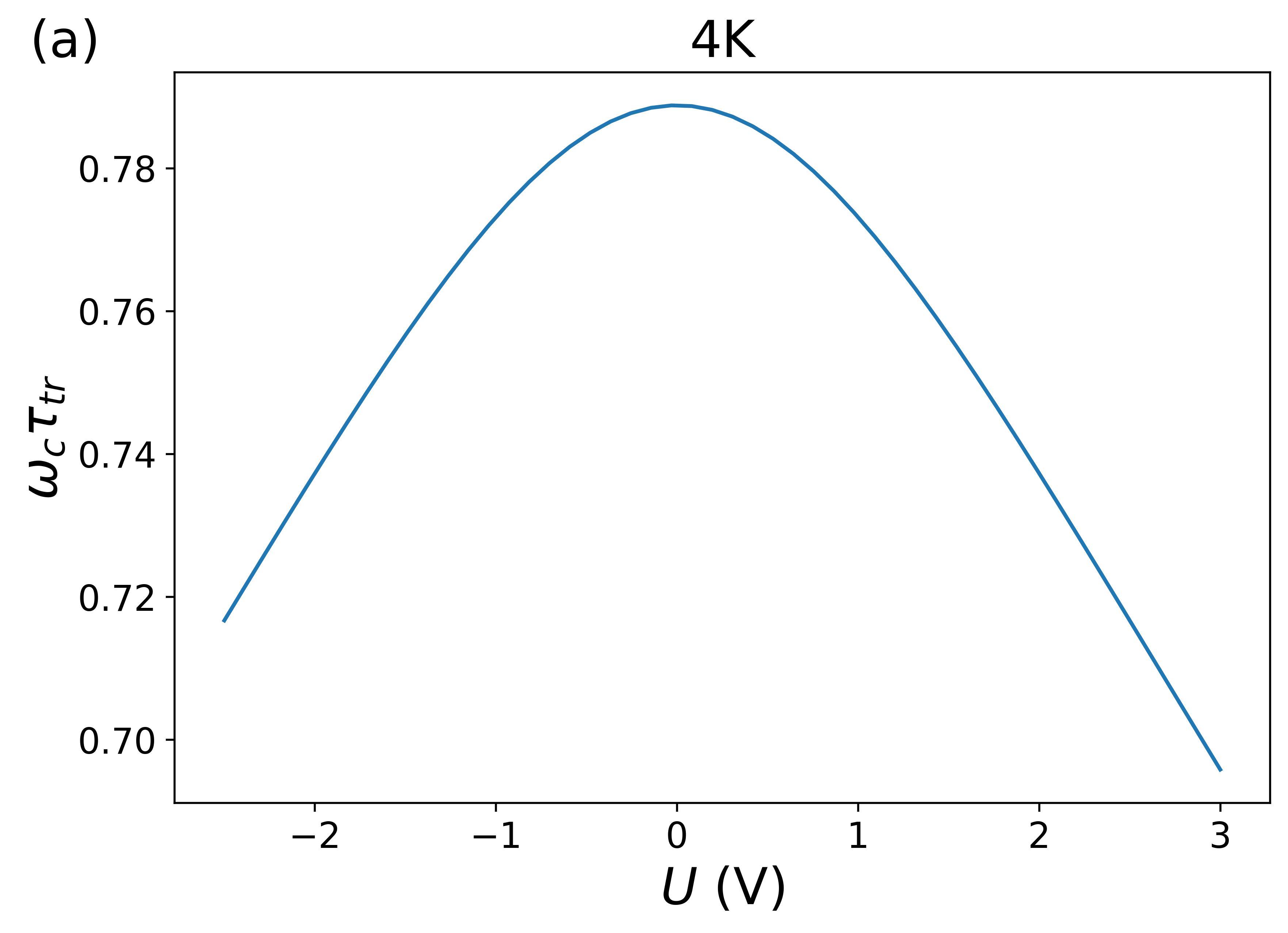}
	\includegraphics[width=.95\linewidth]{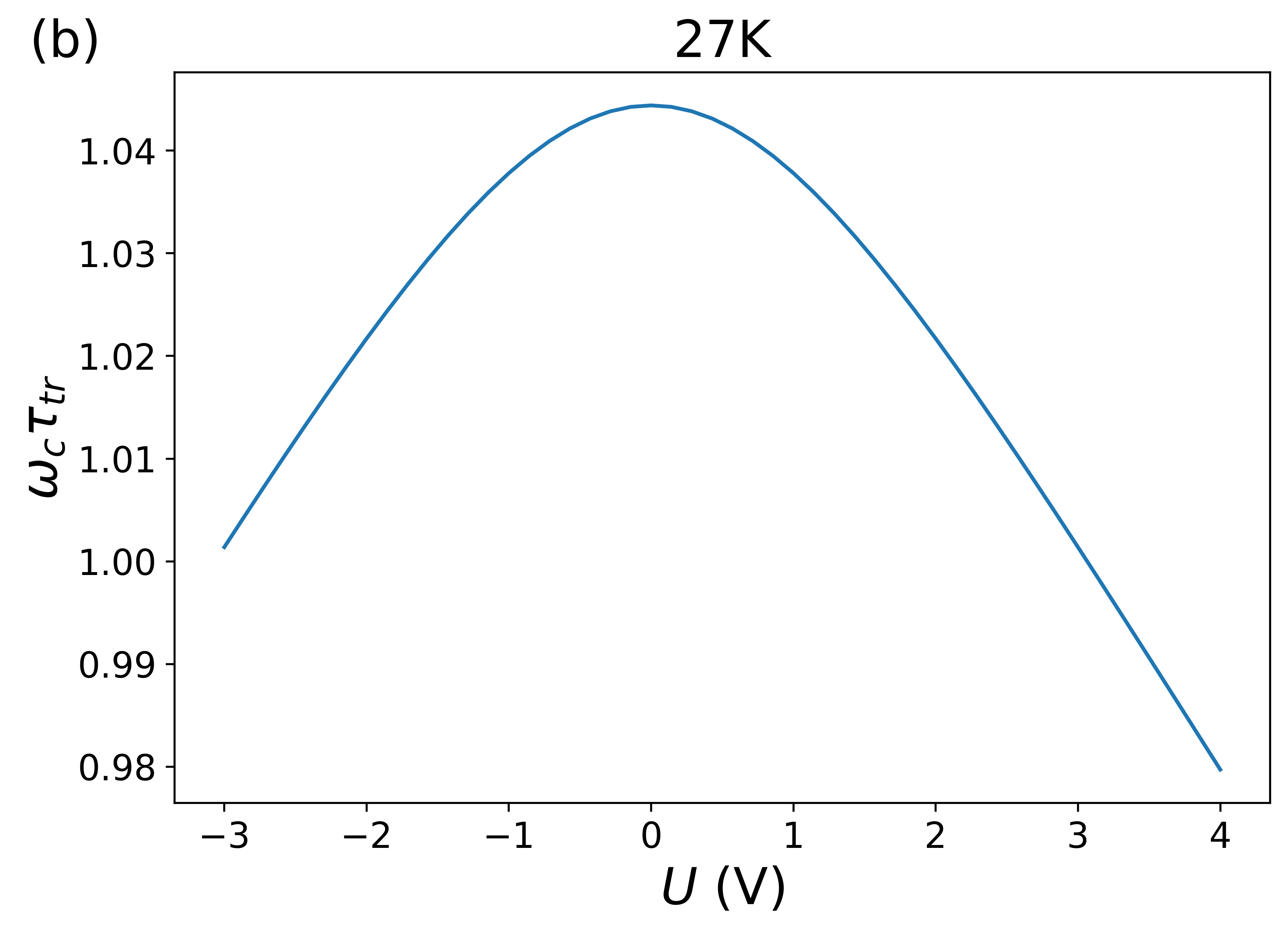}
\caption{The product $\omega_c\tau_\text{tr}$ at $B=0.1$ T with the parameters from the fit (Table \ref{tab:parameters}) for $4$\,K in (a) and $27$\,K in (b). The transport time is larger for the higher temperature, which is consistent with the reduction of the contribution of short-range impurities to scattering processes.    \label{fig:omegaCTau} }
\end{figure}

\section{Details of the fitting procedure}

Here we discuss how the fits are performed and obtain the parameters for Table~\ref{tab:parameters}. The first step is to remove the Dirac point shift as discussed in Appendix~\ref{sec:Diracshift}. In practice, we do not employ the inhomogeneous shift introduced there, but  rather just shift the curves as a whole, such that the maximum is at zero voltage. From the cubic spline through the measured data points, we read off the resistance values at the original voltages and additionally all half integer ones for $27$\,K.

We then fit the obtained magnetoresitances $R(B)-R(0)$ over the whole measured range of $B$ with the fit function Eq.~(\ref{MGres}) for all gate voltages. From the thus obtained parameter $M$ we extract all parameters of the theory according to Eq.~(\ref{eq:theoryParameters}). 
We fit the magnetoresistance with three parameters $\gamma_s$, $c(\alpha)n_\text{imp}^{C}$, and $n_*$,
expressing $M$ as follows:
\begin{align}
\frac{1}{M}=
\pi^2 \hbar \left(n^2+n_*^2\right)
\left[\frac{1}{\gamma_s}+\frac{c(\alpha)n_\text{imp}^{C}}{\sqrt{n^2+n_*^2}}\right].
\end{align}
A plot of the resulting effective disorder parameters $\gamma$ for the two temperatures is shown in Fig. \ref{fig:gammas}.

In Fig.~\ref{fig:omegaCTau}, we also show the product $\omega_c\tau_\text{tr}$ for $B=0.1$T and the two temperatures using the obtained parameters to calculate $\tau_\text{tr}$. 
These plots, demonstrating the dependence of mobility $\mu_0$ on the gate voltage, are in agreement with Fig.~\ref{fig:gammamob}. We see that for $B=0.1$\, T the bending of cyclotron trajectories should be already substantial. Since the quantum scattering time in graphene is smaller by about a factor of two compared to the transport scattering time, the parameter $x=\omega_c\tau_q$ is still smaller than one, and hence Landau levels overlap strong enough to not lead to any more intricate effects in the range magnetic fields $B<0.15$\,T addressed here.

\bibliography{main}

\end{document}